\documentclass[preprint,authoryear,12pt]{elsarticle}

\usepackage{lmodern}
\usepackage{float}
\usepackage{microtype}
\usepackage{xspace}
\usepackage{amsmath,amssymb,amsfonts,amsthm,mathtools}
\usepackage{mathrsfs,bm}
\usepackage[colorlinks=true, linkcolor = blue, hyperfootnotes=false, citecolor = blue]{hyperref}
\usepackage[authoryear]{natbib}
\usepackage{bibentry}
\usepackage[T1]{fontenc}
\usepackage[latin1]{inputenc}
\usepackage[english]{babel}
\usepackage{graphicx}
\usepackage{lineno}
\DeclareMathOperator*{\tr}{tr}
\DeclareMathOperator*{\rank}{rank}
\DeclareMathOperator*{\sign}{sgn}
\newcommand{\norme}[1]{\left\lVert #1 \right\rVert}
\renewcommand{\[}{\begin{equation}}
\renewcommand{\]}{\end{equation}}
\newcommand{\gt}[1]{{\bm{#1}}}
\newcommand{\abs}[1]{\left\lvert #1 \right\rvert}
\newcommand{\scalar}[1]{\left\langle{#1}\right\rangle}
\newcommand{\tp}{\!\otimes\!}
\renewcommand{\t}[1]{{\bf{#1}}}

\renewcommand{\refeq}[1]{(\ref{#1})}

\newcommand{\str}{\varepsilon}
\DeclarePairedDelimiter{\diagfences}{(}{)}
\newcommand{\diag}{\operatorname{diag}\diagfences}
\renewcommand{\phi}{\varphi}

\makeatletter
\def\ps@pprintTitle{%
   \let\@oddhead\@empty
   \let\@evenhead\@empty
   \let\@oddfoot\@empty
   \let\@evenfoot\@oddfoot
}
\makeatother

\begin{document}
\begin{frontmatter}

\title{Microtwist elasticity: A continuum approach to zero modes and topological polarization in Kagome lattices}

\author{Hussein Nassar\footnote{Corresponding author: nassarh@missouri.edu}}
\author{Hui Chen\footnote{Corresponding author: hc6xc@mail.missouri.edu}}
\author{Guoliang Huang\footnote{Corresponding author: huangg@missouri.edu}}

\address{Department of Mechanical and Aerospace Engineering, University of Missouri, Columbia, Missouri 65211, USA}

\begin{abstract}
The topologically polarized isostatic lattices discovered by \citeauthor{Kane2014} (2014, Nat. Phys. 10, 39--45) challenged the standard effective medium theories used in the modeling of many truss-based materials and metamaterials. As a matter of fact, these exhibit Parity (P) asymmetric distributions of zero modes that induce a P-asymmetric elastic behavior, both of which cannot be reproduced within Cauchy elasticity. Here, we propose a new effective medium theory baptized ``microtwist elasticity'' capable of rendering polarization effects on a macroscopic scale. The theory is valid for trusses on the brink of a polarized-unpolarized phase transition in which case they necessarily exhibit more periodic zero modes than they have dimensions. By mapping each periodic zero mode to a macroscopic degree of freedom, the microtwist theory ends up being a kinematically enriched theory. Microtwist elasticity is constructed thanks to leading order two-scale asymptotics and its constitutive and balance equations are derived for a fairly generic isostatic truss: the Kagome lattice. Various numerical and analytical calculations, of the shape and distribution of zero modes, of dispersion diagrams and of polarization effects, systematically show the quality of the proposed effective medium theory. Most notably, the theory is capable of producing a continuum version of Kane and Lubensky's topological polarization vector.
\end{abstract}


\end{frontmatter}
 


\section{Introduction}
Periodic trusses are potent idealized models of several materials such as foams, crystals and metamaterials. When the truss has poor connectivity, the material exhibits a number of zero modes, i.e., deformation modes that cost little to no elastic energy. While catastrophic in many cases, zero modes can still be desirable. In auxetics, for instance, reentrant structures with approximate zero modes provided some of the first examples of materials with negative Poisson's ratio \citep{Lakes1987}. In applications related to smart materials and robotics, non-linear zero modes are essential in structures that can deploy, morph, adapt and move \citep{Milton2013, Milton2013b, Peraza-Hernandez2014, Khanikaev2015, Rocklin2015a, Nassar2017a, Nassar2018b, Rus2018a}. But perhaps the most spectacular application of zero modes in recent years has been in the design of acoustic ``invisibility'' cloaks. Indeed, form-invariance, a cornerstone of transformation-based cloaking, can only be fulfilled thanks to materials with a number of non-trivial zero modes. In acoustics, \cite{Norris2008} identified these materials to be \citeauthor{Milton1995}'s (\citeyear{Milton1995}) pentamodes; in full elasticity, other materials with zero modes are just as useful \citep{Nassar2018a, Nassar2019, Nassar2020, Xu2020}.

From the point of view of the material's constitutive law $\gt\sigma=\t C^*:\gt\str$, zero modes appear when the effective elasticity tensor $\t C^*$ is singular. Thus, zero modes correspond to compatible fields of strain $\gt\str_o(\t x)$ such that $\t C^*:\gt\str_o=\t 0$ at each position $\t x$. Remarkably, if $\gt\str_o(\t x)$ is a zero mode then so is $\gt\str_o(-\t x)$. More generally, Parity (P) symmetry, namely the invariance of the set of solutions under the spatial inversion $\t x\mapsto - \t x$, is a key feature of Cauchy's theory of elasticity. Nonetheless, there are trusses where zero modes systematically grow in amplitude in a preferential direction and systematically decay in the opposite direction \citep{Lubensky2015a, Mao2018}. Materials with such underlying trusses have a broken P-symmetry; we say that they are polarized. Other trusses admit zero modes for which $\gt\str_o(\t x)=\t 0$ (see, e.g., the same references). To capture such zero modes on the level of the material requires finer measures of strain besides $\gt\str$ and its gradients. In both cases, Cauchy's theory is unsatisfactory. It is the purpose of the present paper to propose an enriched effective medium theory capable of faithfully reproducing microstructural zero modes and related polarization effects on the continuum scale. Derivations are carried for a fairly generic truss: the Kagome lattice.

Polarized Kagome lattices came to our attention while reading the elegant work of \cite{Kane2014} on topological polarization in isostatic lattices. In the detail, a regular, e.g. the standard, Kagome lattice exhibits bulk zero modes which maintain uniform amplitude across the whole truss. General geometric distortions of the lattice then gap these modes at non-zero wavenumbers in the same way certain frequencies are gapped in a phononic crystal except that the gapped frequency is the zero frequency. Hence, zero modes become ``evanescent''; they adopt exponential profiles that decay towards the bulk and re-localize at free boundaries. Kane and Lubensky characterized the conditions under which the re-localization of zero modes towards free boundaries happens unevenly and favors certain boundaries over their opposites. Note that the found conditions and the resulting P-asymmetric distribution of zero modes are topological in nature, i.e., they are immune to continuous perturbations, small and large, so long as the aforementioned zero-frequency gap remains open. This is why such Kagome lattices are qualified as ``topologically polarized''. Based on these principles, \citet{Bilal2017} designed and tested a material featuring a polarized elastic behavior. A finite slab of their material appears soft when indented on one side and hard when indented on the opposite side. Elastic polarization effects are not restricted to boundaries and are expected to emerge in the bulk as well; see, e.g., \cite{Rocklin2017}.

Our aim therefore is to reconcile the above observations with an effective theory of elasticity. Following asymptotic analysis, we find that the theory naturally maps the periodic zero modes of the truss to macroscopic Degrees Of Freedom (DOFs). For instance, regular Kagome lattices admit three periodic zero modes, two translations and the so-called periodic twisting. While translations are mapped to the macroscopic displacement field $\t U$, periodic twisting is mapped to an extra DOF $\phi$. The resulting effective continuum is called the ``microtwist'' continuum after the additional periodic zero mode. The microtwist continuum also has two extra measures of strain, $\phi$ itself and its gradient, and by way of duality, two extra measures of stress. By continuity, nearly-regular or weakly-distorted Kagome lattices are also described in the same way albeit with different effective properties. In that case, periodic twisting is no longer a zero mode strictly speaking but still corresponds to a highly compliant mechanism. By contrast, we do not deal with strongly-distorted lattices: these may exhibit strong polarization effects but only within thin boundary layers. We speculate that Cauchy elasticity with ad-hoc boundary or jump conditions is satisfactory for their continuum modeling; see, e.g., the papers by \cite{Marigo2016, Marigo2017}.

Microtwist elasticity is the outcome of leading order two-scale asymptotic expansions. It is reminiscent of $\t k\cdot\t p$ perturbation theory used in condensed matter physics \citep{Dresselhaus2008}. In that language, the theory describes the asymptotic behavior of Kagome lattices near the $\Gamma$ point when the acoustic branches and the first optical branch are strongly coupled, i.e., degenerate or nearly degenerate. Furthermore, the theory bears resemblance to high-frequency asymptotic homogenization theories (see, e.g., \citealp{Bensoussan1978,Craster2010, Allaire2011, Harutyunyan2016}).

Several earlier contributions sought generalized effective media for trusses, be them of the micropolar type \citep{Lakes1982, Lakes2001, Spadoni2012, Liu2012a, Bacigalupo2014, Chen2014b, Frenzel2017} or the strain gradient type \citep{Auffray2010, Bacigalupo2014, Rosi2016}. Often, the aim was to model chiral effects. In that regard, it is worth stressing that chirality, or anisotropy of any kind for that matter, is fundamentally different from P-asymmetry. Indeed, when the former is concerned with the action of rotations on the constitutive law, the latter is concerned with the action of the inversion $\t x\mapsto -\t x$ on fields solution to the motion equation. See, e.g., \cite{Nassar2020} for a theory of elasticity that is chiral but P-symmetric. More relevant to our purposes is the work of \cite{Sun2012} who hinted at microtwist elasticity in a particular case but did not pursue a full theory. More recently, \cite{Sun2019} and \cite{Saremi2020} proposed theories for polarized effective media of the strain gradient type. Our asymptotic analysis suggests that a kinematically enriched medium is indispensable, at least in the strong coupling limit of interest here.

The paper goes as follows: in Section~\ref{sec:geo}, we classify general Kagome lattices in two phases, regular and distorted, based on a count of their periodic and Floquet-Bloch zero modes. In Section~\ref{sec:Homo}, we argue why enriching the effective medium is necessary in the case of regular and weakly-distorted lattices. Subsequently, we deploy two-scale asymptotics and deduce, in closed form, the constitutive and balance equations governing the effective microtwist continuum. In Section~\ref{sec:SPMC}, we evaluate how well the microtwist theory can predict zero modes, polarization effects and dispersion relations at zero and finite frequencies both qualitatively and quantitatively. Most importantly, we demonstrate how the microtwist theory provides an accurate continuum version of the topological polarization of \cite{Kane2014}.
\section{Kagome lattices and their zero modes}\label{sec:geo}
General Kagome lattices are introduced and classified into two phases, regular and distorted, based on the number and type of zero modes they support. The analysis here is based on the discrete lattice model. A continuum model, suitable for regular and weakly-distorted lattices, will be derived in the next section.
\subsection{Kinematics and dynamics of Kagome lattices}
\begin{figure}[ht]
\centering
\includegraphics[keepaspectratio,width=\textwidth]{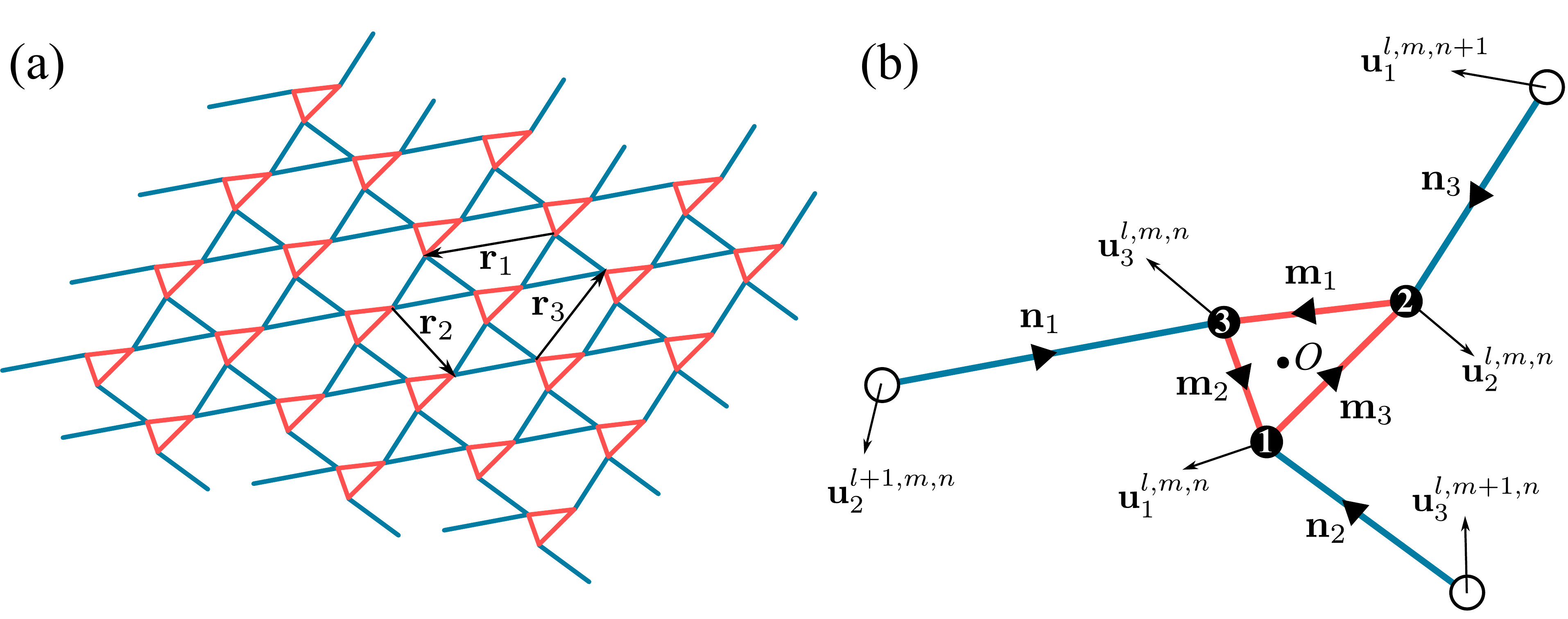}
\caption{A general Kagome lattice: (a) a periodic reference configuration; (b) a magnified and annotated unit cell. The displacements of the nodes are shown as arrows. The solid and empty circles represent the interior and exterior nodes of the unit cell respectively. Blue bonds with unit vectors $\t m_j$ and red bonds with unit vectors $\t n_j$ have respective lengths $a_j$ and $b_j$ and respective spring constants $\alpha_j$ and $\beta_j$.}
\label{fig:fig1}
\end{figure}

Consider the general Kagome lattice depicted in Figure~\ref{fig:fig1}a in a periodic reference configuration. The lattice is made of a set of massless spring-like edges connecting massive hinge-like nodes. Vectors $\t r_j$ are lattice vectors: the reference configuration is invariant by translation along any integer linear combination of the $\t r_j$. A unit cell is shown on Figure~\ref{fig:fig1}b: it has three nodes in its interior, i.e., the filled circles, indexed with $j\in\{1,2,3\}$ and initially placed at $\t x_j$. Index $j$ is always understood modulo $3$: if $j=3$ then $j+1=1$ and if $j=1$ then $j-1=3$. Exterior to the unit cell, but at its boundary, there are three other nodes drawn as empty circles and whose initial positions are given by $\t x_j+\t r_{j-1}$.  Thus, the initial positions of all nodes can be deduced from the $\t x_j$ according to
\[
\t x_j^{l,m,n} = \t x_j + \t x^{l,m,n},\quad \t x^{l,m,n} = l\t r_1 + m \t r_2 + n \t r_3,\quad (l,m,n)\in\mathbb{Z}^3.
\]
Here, $\t x_j^{l,m,n}$ designates the position of node $j$ of unit cell $(l,m,n)$. The use of three indices, $l$, $m$ and $n$, to describe a 2D lattice may seem superfluous. Indeed, one has $\t r_1+\t r_2+\t r_3=\t 0$ and any combination of $\t r_1$, $\t r_2$ and $\t r_3$ can be reduced to one where, say, only $\t r_1$ and $\t r_2$ are present. Nonetheless, in order to enforce the formal permutation symmetry, namely that the nodes within a unit cell play equivalent roles and can be numbered arbitrarily, it is preferable to maintain the use of three vectors $\t r_j$ without expanding any one along the other two. This attitude will greatly simplify later derivations. Note that, as a side effect, the coordinates $(l,m,n)$ of a unit cell are not unique. For instance, $(0,0,0)$ and $(1,1,1)$ designate the same unit cell. If uniqueness is desired, then one can require the satisfaction of some constraint such as $0\leq l+m+n\leq 2$ but this will not be enforced and should have no influence on what follows.

A unit cell further has six edges oriented along the unit vectors $\t m_j$ (red bonds) and $\t n_j$ (blue bonds) and of respective lengths $a_j$ and $b_j$; see Figure~\ref{fig:fig1}b. The Kagome lattice is therefore isostatic in the sense that it has as much DOFs as it has bonds per unit cell: six of each. The displacement of node $j$ in unit cell $(l,m,n)$ is called $\t u_{j}^{l,m,n}$ whereas the elongation of edge $\t m_j$ (resp. $\t n_j$) is called $y_j^{l,m,n}$ (resp. $z_j^{l,m,n}$). These are related through
\[\label{eq:Deltas}
\begin{aligned}
y_j^{l,m,n} &= \scalar{\t m_j,\t u_{j-1}^{l,m,n}-\t u_{j+1}^{l,m,n}},\quad
&&z_1^{l,m,n} = \scalar{\t n_1,\t u_{3}^{l,m,n}-\t u_{2}^{l+1,m,n}},\\
z_2^{l,m,n} &= \scalar{\t n_2,\t u_{1}^{l,m,n}-\t u_{3}^{l,m+1,n}},\quad
&&z_3^{l,m,n} = \scalar{\t n_3,\t u_{2}^{l,m,n}-\t u_{1}^{l,m,n+1}}.
\end{aligned}
\]
The tensions in the corresponding edges are given by $\alpha_jy_j^{l,m,n}$ and $\beta_jz_j^{l,m,n}$ where $\alpha_j$ and $\beta_j$ are the spring constants of edges $\t m_j$ and $\t n_j$ respectively. Thus, the internal force $\t t_j^{l,m,n}$ acting on node $j$ in unit cell $(l,m,n)$ reads
\[\label{eq:tensions}
\begin{split}
\t t_1^{l,m,n} &= 
-\alpha_2 y_2^{l,m,n}\t m_{2}
-\beta_2 z_2^{l,m,n}\t n_{2}
+\alpha_3 y_3^{l,m,n}\t m_{3}
+\beta_3 z_3^{l,m,n-1}\t n_{3},\\
\t t_2^{l,m,n} &=
-\alpha_3y_3^{l,m,n}\t m_{3}
-\beta_3z_3^{l,m,n}\t n_{3}
+\alpha_1y_1^{l,m,n}\t m_{1}
+\beta_1z_1^{l-1,m,n}\t n_{1},\\
\t t_3^{l,m,n} &=
-\alpha_1y_1^{l,m,n}\t m_{1}
-\beta_1z_1^{l,m,n}\t n_{1}
+\alpha_2y_2^{l,m,n}\t m_{2}
+\beta_2z_2^{l,m-1,n}\t n_{2}.
\end{split}
\]

Finally, Newton's second law can be stated as
\[\label{eq:NSL}
\t t^{l,m,n}_j + \t f^{l,m,n}_j = m_j\ddot{\t u}^{l,m,n}_j,
\]
where $m_j$ is the mass of node $j$ and $\t f^{l,m,n}_j$ is an external force applied to node $j$ of unit cell $(l,m,n)$.

In what follows, without loss of generality, we let the origin of coordinates ``$O$'' be the geometric center of the red triangle $\Delta \equiv (a_1 \t m_1,a_2 \t m_2,a_3 \t m_3)$. Accordingly, the reference positions of the three interior nodes, with respect to the origin, are
\[
\t x_j = \frac{a_{j+1}\t m_{j+1}-a_{j-1}\t m_{j-1}}{3}.
\]
For later purposes, we also define $\bar{\t x}_j$ to be the image of $\t x_j$ by a plane rotation of angle $\pi/2$. More generally, a superimposed bar will symbolize a plane rotation of $\pi/2$.
\subsection{Zero modes}
We call \emph{zero mode}, a static solution to Newton's equation in the absence of external loading, i.e., a solution $\t u_j^{l,m,n}$ to
\[
\t t^{l,m,n}_j = \t 0.
\]
Equivalently, a zero mode is a configuration of the lattice which stretches and compresses no bonds so that
\[
y_j^{l,m,n}=z_j^{l,m,n}=0.
\]

\begin{figure}[ht]
\centering
\includegraphics[keepaspectratio,width=0.5\textwidth]{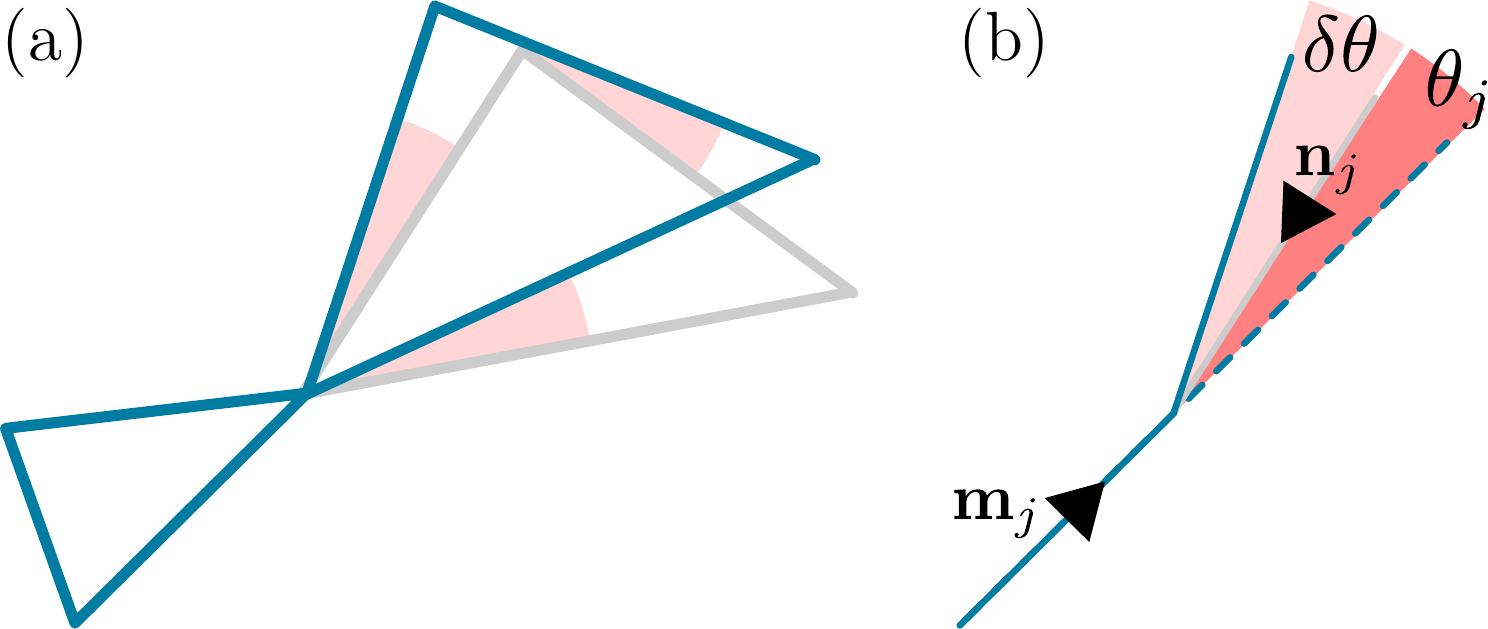}
\caption{Elementary twisting mechanism (a) and corresponding notations (b). Edges are in gray in their initial positions and colored in their current positions.}
\label{fig:mech}
\end{figure}

In this sense, rigid body translations and rotations are zero modes. Kagome lattices admit a number of other, more interesting, zero modes all inherited from the elementary \emph{twisting} mechanism illustrated on Figure~\ref{fig:mech}. Understanding the zero modes of Kagome lattices is essential to justify the need for the generalized theory of elasticity introduced in the next section. Thus, zero modes are studied in the remainder of this section in some detail. This is also an occasion to gain insight into the geometry of Kagome lattices, to familiarize the reader with the introduced notations and to improve a number of results obtained by other authors. In particular, we will investigate periodic and Floquet-Bloch zero modes.
\subsection{Periodic zero modes}\label{subsec:FZMZ}
We call \emph{periodic} a configuration that does not depend on the indices $(l,m,n)$ of unit cells, i.e.,
\[
\t u^{l,m,n}_j=\t u_j.
\]
Zero mode or not, dismissing the dependence over $(l,m,n)$ greatly simplifies the governing equations. For instance, elongations are given by the matrix product
\[\label{eq:C1}
\begin{bmatrix}
y_1\\
y_2\\
y_3\\
z_1\\
z_2\\
z_3
\end{bmatrix}
=
C
\begin{bmatrix}
\t u_1\\\t u_2\\\t u_3
\end{bmatrix},\quad
C =
\begin{bmatrix}
\t 0 & -\t m_1' & \t m_1'\\
\t m_2' & \t 0 & -\t m_2'\\
-\t m_3' & \t m_3' & \t 0\\
\t 0 & -\t n_1' & \t n_1'\\
\t n_2' & \t 0 & -\t n_2'\\
-\t n_3' & \t n_3' & \t 0
\end{bmatrix},
\]
where $C$ is a $6\times 6$ compatibility matrix and a prime means conjugate transpose so that $\t m'_j\t u_k = \scalar{\t m_j,\t u_k}$. Accordingly, a periodic zero mode solves
\[
C\Phi = 0, \quad
\Phi =  \begin{bmatrix}
\t u_1\\\t u_2\\\t u_3
\end{bmatrix}.
\]
Hence, periodic zero modes are null vectors of matrix $C$. By the rank-nullity theorem \citep{Birkhoff1998}, their number is equal to $Z = 6-\rank C$ where $6$ is the dimension of $C$ and $\rank C$ its rank.

\begin{figure}[ht!]
\centering
\includegraphics[keepaspectratio,width=\textwidth]{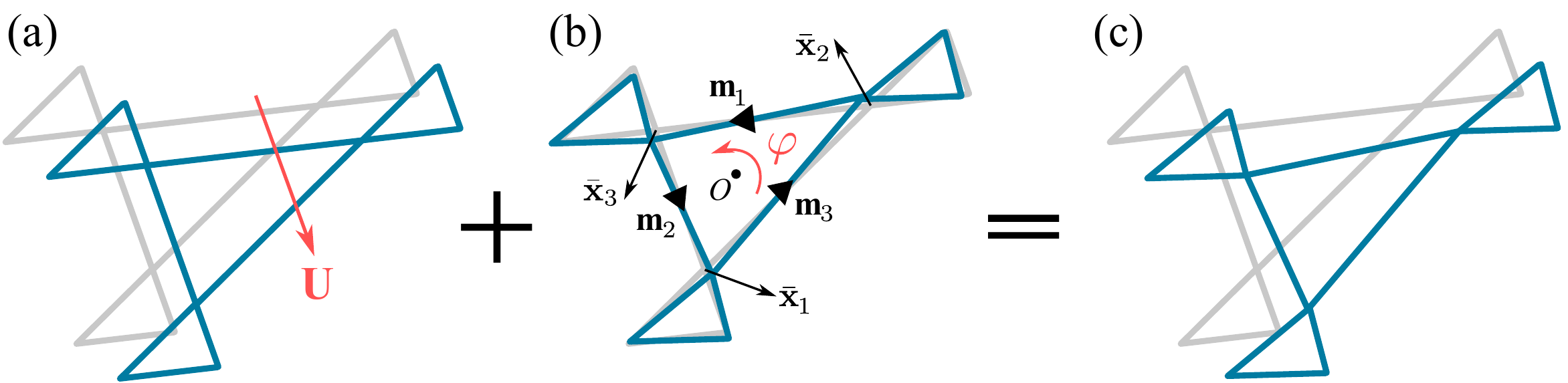}
\caption{Generic periodic zero mode of a regular Kagome lattice: (a) global translation $\t U$; (b) a twisting motion of angle $\varphi$ around the center $O$; (c) a linear combination of a translation and a twisting motion. The initial and deformed configurations are traced in grey and blue respectively.}
\label{fig:fig2}
\end{figure}

Translations by a vector $\t U$ are characterized by $\t u_1=\t u_2=\t u_3=\t U$ (Figure~\ref{fig:fig2}a). They take the form
\[\label{eq:D}
\Phi = \begin{bmatrix}
\t U\\\t U\\\t U
\end{bmatrix}
= D\t U, \quad
D =
\begin{bmatrix}
\t I\\\t I\\\t I
\end{bmatrix},
\]
where $\t I$ is the second-order identity tensor. These clearly satisfy $C\Phi = 0$. Translations span two periodic zero modes. It is not too hard to show that if $\theta_j\neq 0$, for some $j$, then $\rank C = 4$ and $Z=2$. Such lattices will be called \emph{distorted}: they admit no other periodic zero modes besides translations (Appendix A). Otherwise, if $\theta_j=0$, for all $j$, then $\rank C=3$ and $Z=3$. Such lattices will be called \emph{regular}. These admit one extra periodic zero mode given by the twisting motion
\[\label{eq:T}
\Phi = 
\begin{bmatrix}
\bar{\t x}_1\\
\bar{\t x}_2\\
\bar{\t x}_3
\end{bmatrix} \phi
 \equiv T\phi.
\]

Restricted to the nodes of one unit cell, twisting is a rotation whose center can be chosen arbitrarily. Here, the geometric center ``$O$'' is chosen as the center of rotation whereas $\phi$ is the angle of rotation (Figure~\ref{fig:fig2}b). It is easy to check that $T$ is indeed a zero mode, i.e., that $CT = 0$. While doing so it is useful to verify first that $\bar{\t x}_{j-1}-\bar{\t x}_{j+1} = a_j\bar{\t m}_j$ is orthogonal to both $\t m_j$ and $\t n_j$, these two being parallel in regular lattices.

In conclusion, the periodic zero modes of a regular Kagome lattice are given by the linear combination of translations and a twisting motion (Figure~\ref{fig:fig2}c)
\[
\Phi = D\t U + T\phi,
\]
or equivalently by
\[
\t u_j = \t U + \phi\bar{\t x}_j.
\]
\subsection{Floquet-Bloch zero modes}
\emph{Floquet-Bloch zero modes} take the form
\[
\t u^{l,m,n}_j = \t u_j \exp\left(i\scalar{\t q,\t x^{l,m,n}}\right)
\]
where $\t q$ is a real wavenumber. Alternatively, we can write
\[
\t u^{l,m,n}_j=P_1^lP_2^mP_3^n\t u_j,
\]
where the $P_j$ are unitary complex numbers such that $P_1P_2P_3=1$. We can pass from one form to the other by setting $q_j=\scalar{\t q,\t r_j}$ and $P_j=e^{iq_j}$. Elongations admit similar expressions
\[
y_j^{l,m,n} = P_1^lP_2^mP_3^n y_j,\quad
z_j^{l,m,n} = P_1^lP_2^mP_3^n z_j,
\]
and it is again convenient to introduce a compatibility matrix as in~\refeq{eq:C1} but with $C$ replaced by
\[
C(\t q) =
\begin{bmatrix}
\t 0 & -\t m_1' & \t m_1'\\
\t m_2' & \t 0 & -\t m_2'\\
-\t m_3' & \t m_3' & \t 0\\
\t 0 & -P_1\t n_1' & \t n_1'\\
\t n_2' & \t 0 & -P_2\t n_2'\\
-P_3\t n_3' & \t n_3' & \t 0
\end{bmatrix}.
\]
Then, Floquet-Bloch zero modes of wavenumber $\t q$ exist if and only if the linear system
\[
C(\t q)\Phi = \t 0,\quad
\Phi=\begin{bmatrix}
\t u_1\\\t u_2\\\t u_3
\end{bmatrix},
\]
of six equations has a non-trivial solution. The first three equations are automatically satisfied by the ansatz
\[
\t u_j = \t U + \phi \bar{\t x}_j,
\]
where $\t U$ and $\phi$ are the new unknowns. The remaining three equations become
\[\label{eq:PZMbis}
\begin{bmatrix}
(1-P_1)\t n_1' & \scalar{\t n_1,\bar{\t x}_3 - P_1\bar{\t x}_2}\\
(1-P_2)\t n_2' & \scalar{\t n_2,\bar{\t x}_1 - P_2\bar{\t x}_3}\\
(1-P_3)\t n_3' & \scalar{\t n_3,\bar{\t x}_2 - P_3\bar{\t x}_1}
\end{bmatrix}
\begin{bmatrix}
\t U \\ \phi
\end{bmatrix} =
\begin{bmatrix}
0 \\ 0 \\ 0
\end{bmatrix}
\]
and have a zero-determinant condition equivalent to
\[\label{eq:PZM}
\prod_j(1-P_j)\sum_j \scalar{b_j\t n_j,\bar{\t x}_{j+1}} +
\sum_j (1-P_{j-1})(1-P_{j+1})\scalar{b_j\t n_j,a_j\bar{\t m}_j} = 0.
\]
From the above equation, it is clear that $\t q = \t 0$, $P_1=P_2=P_3=1$ provide systematic solutions. These are no other than the periodic zero modes of the previous subsection. For $\t q\neq\t 0$, it can be verified that $P_j=1$ (i.e., $\t q\perp\t r _j$) is a solution if and only if $\t m_j$ and $\t n_j$ are colinear. No other solutions exist as long as the lattice is not ``too distorted'' (see Appendix B).
\begin{figure}[ht!]
\centering
\includegraphics[keepaspectratio,width=\textwidth]{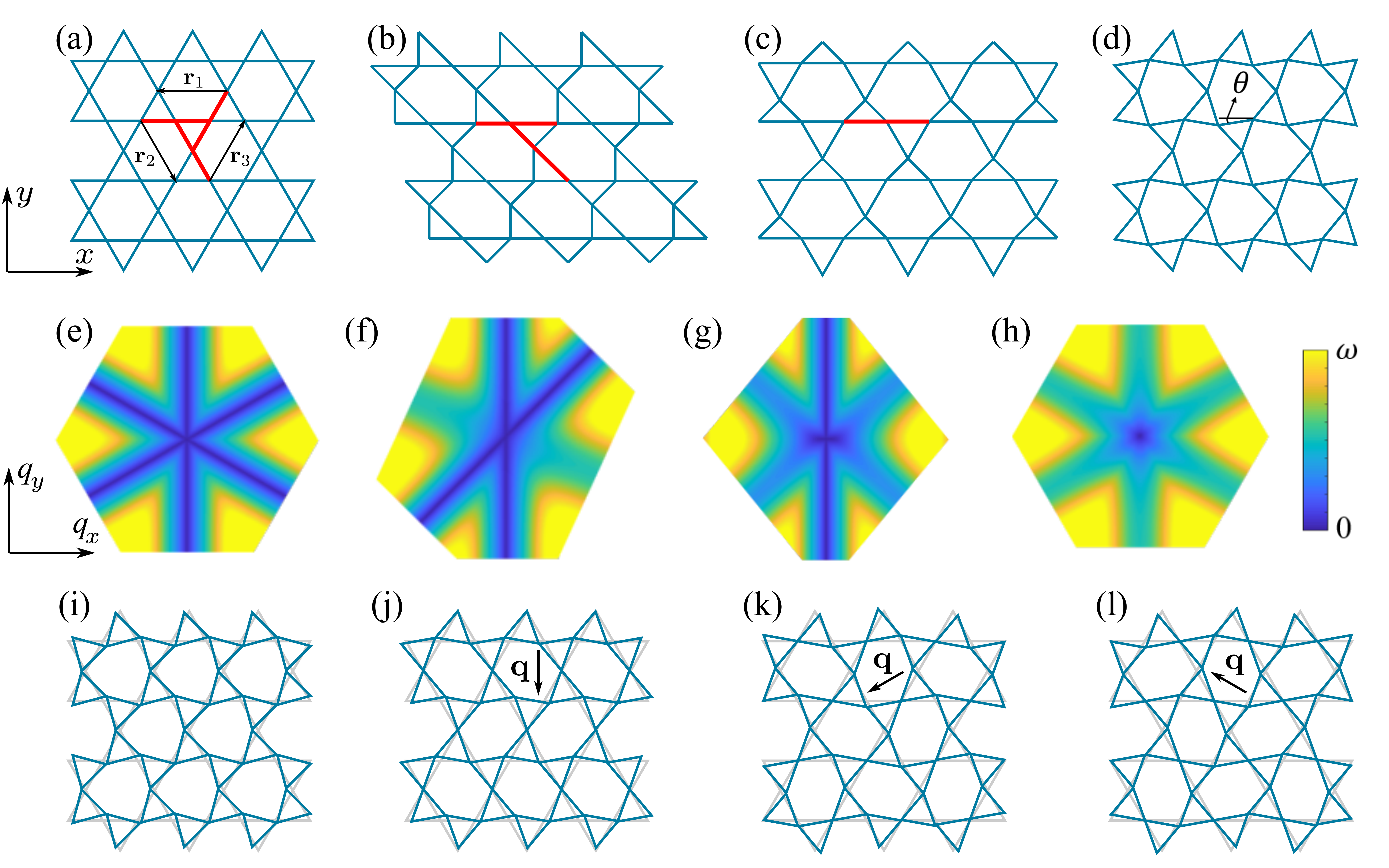}
\caption{Examples of regular (a) and distorted Kagome lattices (b-d) and of their low isofrequency contours (e-h). Red solid lines correspond to pairs of colinear bonds (i.e., $\t m_j = - \t n_j$). Examples of Floquet-Bloch zero modes acting on the regular lattice (a) are shown in (i-l): (i) periodic zero mode ($\t q = \t 0$); (j-l) Floquet-Bloch zero modes with $\t q\perp\t r_j$. Numerical parameters: $\alpha_j = \beta_j = 1$; in (a), $a_j = b_j = 1$; in (b), $(a_1,a_2,a_3,b_1,b_2,b_3) = (\sqrt{2},1,1,1,\sqrt{2},1)$; in (c), $(a_1,a_2,a_3,b_1,b_2,b_3) = (1,1,1,1,1/\sqrt{2},1/\sqrt{2})$; (d) is (a) twisted through an angle $\theta = 5^{\circ}$.}
\label{fig:fig3}
\end{figure}

It is then possible to draw the locus of real wavenumbers $\t q$ for which Floquet-Bloch zero modes exist. Such a ``zero-frequency dispersion diagram'' is composed of a number $n$ of straight lines where $n$ is the number of colinear pairs $(\t m_j,\t n_j)$. As an illustration, we plot the isofrequency contours of example regular and distorted Kagome lattices within an arbitrary low-frequency range $(0, \omega)$. The regular lattice (Figure~\ref{fig:fig3}a) has three pairs of colinear bonds highlighted in red; thus, its spectrum (Figure~\ref{fig:fig3}e) shows Floquet-Bloch zero modes in the three directions perpendicular to the three lattice vectors $\t r_j$; these appear in the deep blue color corresponding to zero frequency. Figure~\ref{fig:fig3}i shows the mode shape of a periodic zero mode ($\t q = 0$) and Figures~\ref{fig:fig3}j-\ref{fig:fig3}l show mode shapes of zero modes with wavenumbers in the directions $\bar{\t r}_j$. Subsequently, distortions that break the alignment of one, two or three pairs of bonds gap one, two or three lines of zero modes, respectively. The resulting lattices and spectra are depicted in Figure~\ref{fig:fig3}b-d and f-h.
\section{Homogenization of Kagome lattices: the microtwist continuum}\label{sec:Homo}
\subsection{Prelude: Why Cauchy elasticity is not enough}
Having explored Kagome lattices from a purely geometric point of view, it is time to investigate their elastic behavior. We are particularly concerned here with the homogenization limit, i.e., the limit of infinitesimal unit cells.
The standard theory of elasticity then permits to model a Kagome lattice as a homogeneous Cauchy continuum where reigns a stress distribution related to a strain field through Hooke's law
\[
\gt\sigma = \t C^*:\t \nabla^s\t U,
\]
with $\t C^*$ being the homogenized fourth-order tensor of effective elastic moduli. Interestingly, the panorama of zero modes presented above helps appreciate why Cauchy's continuum is bound to fail in its task for regular lattices. On one hand, Cauchy's continuum accounts exactly for $d$ acoustic branches in $d$ dimensions. However, the presence of three periodic zero modes in regular Kagome lattices means that they exhibit three acoustic branches, or two acoustic branches degenerate with one optical branch. On the other hand, the zero-frequency dispersion diagram of Cauchy's continuum usually consists of a single point but generally can comprise up to $d$ straight lines in $d$ dimensions. Furthermore, in that case, tensor $\t C^*$ is necessarily singular and has either a vanishing bulk modulus or a vanishing shear modulus. In contrast, regular Kagome lattices have three straight lines in their zero-frequency dispersion diagram and no vanishing elastic moduli. Note that strain gradient media more generally suffer from the same limitations.

Extrapolating by continuity, we argue that Cauchy's continuum will not be a poor model only for regular Kagome lattices but also for weakly-distorted ones, i.e., lattices where the $\theta_j$ do not vanish necessarily but remain in any case close to zero. This motivates our speaking of a phase transition: we call a \emph{phase transition} any perturbation, geometric or otherwise, that transforms a regular lattice into a distorted one thus changing the number of supported periodic zero modes from $3$ to $2$. Lattices on the brink of a phase transition are therefore regular or weakly-distorted lattices for which the inequality $\abs{\theta_j}\ll\pi$ holds for all $j$. These lattices, based on our discussion, necessitate a richer continuum than that of Cauchy for their accurate modeling. In the remainder of this section, we find that continuum.
\subsection{Three perturbations}
Starting with a regular Kagome lattice, we introduce three perturbations.

\begin{figure}[ht!]
\centering
\includegraphics[scale=1]{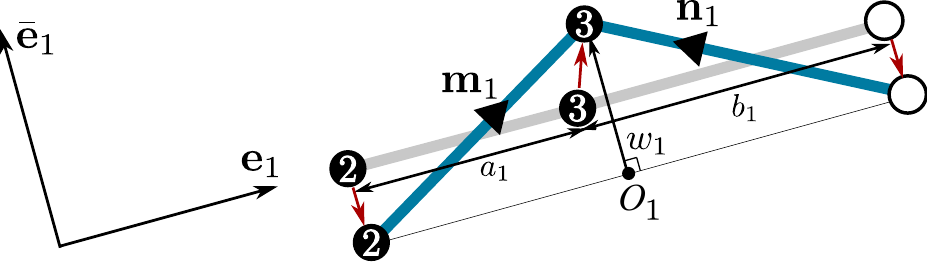}
\caption{Illustration of a distortion inducing a phase transition (red arrows): The unmarked node belongs to the boundary of the unit cell and is displaced by the same vector as node $2$ so as to maintain periodicity. Elevation $w_1$ is positive when the vector running from point $O_1$ to node $3$ is in the same direction as $\bar{\t e}_1$. The elevations $w_j$ are the only parameters relevant here.}
\label{fig:fig5}
\end{figure}

First, we induce a phase transition by perturbing the initial positions of the nodes so as to break the alignment of any one of the three pairs $(\t m_j,\t n_j)$; see Figure~\ref{fig:fig5}. Letting $(\t e_j,\bar{\t e}_j)$ be an orthonormal basis colinear to $(\t r_j,\bar{\t r}_j)$, a nearly-regular lattice is obtained and is characterized by
\[\label{eq:EFExp}
\t m_j = \t e_j + \frac{w_j}{a_j} \bar{\t e}_j + \dots,\quad
\t n_j = -\t e_j + \frac{w_j}{b_j} \bar{\t e}_j + \dots,
\quad \abs{w_j}\ll \min(a_j,b_j).
\]
Above, the parameters $w_j$ control the geometric distortion and are related to the angles $\theta_j$ through $\theta_j\approx -w_j(1/a_j + 1/b_j)$. Thus, the perturbation leads to a distorted lattice if at least one $w_j$ is non-zero.

Second, we no longer assume the displacements $\t u_j^{l,m,n}$ to be periodic, or linear. Instead, we let them derive from slowly varying smooth functions $\t u_j(\t x)$ upon replacing $\t x$ with $\t x^{l,m,n}$. That is $\t u_j(\t x^{l,m,n}) = \t u_j^{l,m,n}$. More importantly, the leading-order Taylor expansions
\[\label{eq:ansatz}
\begin{split}
\t u_j^{l+1,m,n} - \t u_j^{l,m,n}&= \partial_1\t u_j,\\
\t u_j^{l,m+1,n} - \t u_j^{l,m,n}&= \partial_2\t u_j,\\
\t u_j^{l,m,n+1} - \t u_j^{l,m,n}&= \partial_3\t u_j,
\end{split}
\]
hold with $\partial_j = \scalar{\t r_j,\gt\nabla}$ being the differential with respect to $\t x$ in direction~$\t r_j$. Then, the functions $\t u_j$ are slowly varying in space if and only if $\norme{\partial_j}\ll 1$.

Third, we no longer assume the displacements $\t u_j^{l,m,n}$ to be static but allow them to change with respect to time at small rates $\omega$ satisfying $\omega \sqrt{\max(m_j)}\ll \sqrt{\min(\alpha_j,\beta_j)}$.

Accordingly, in what follows, the behavior of Kagome lattices is investigated in the homogenization limit and, specifically, in the critical regime
\[\label{eq:regime}
\norme{\partial_j} \sim \sqrt{\frac{\max(m_j)}{\min(\alpha_j,\beta_j)}}\omega \sim
\frac{\abs{w_j}}{\min(a_j,b_j)} \ll 1
\]
where all three introduced perturbations are \emph{a priori} of the same order of magnitude.
\subsection{Asymptotic expansions}
We start by revisiting the equations of the previous section and replace them with their second-order asymptotic expansions. For instance, injecting~\refeq{eq:EFExp} and~\refeq{eq:ansatz} back into~\refeq{eq:Deltas} yields
\[
\begin{bmatrix}
y_1\\
y_2\\
y_3\\
z_1\\
z_2\\
z_3
\end{bmatrix}
=
C
\begin{bmatrix}
\t u_1\\\t u_2\\\t u_3
\end{bmatrix}
=C\Phi,
\quad
C = C_0 +\delta C + \delta^2C + \dots,
\]
where the elongations $y_j$ and $z_j$ and displacements $\t u_j$ are all functions of the space variable $\t x$; $C$ is a differential compatibility operator; $C_0$ is its restriction to static periodic configurations over a regular lattice; and $\delta C$ and $\delta^2C$ are its first-order and second-order corrections. We have previously encountered $C_0$ in equation~\refeq{eq:C1}; here it specifies into
\[\label{eq:comp0}
C_0 =
\begin{bmatrix}
\t 0 & -\t e_1' & \t e_1'\\
\t e_2' & \t 0 & -\t e_2'\\
-\t e_3' & \t e_3' & \t 0\\
\t 0 & \t e_1' & -\t e_1'\\
-\t e_2' & \t 0 & \t e_2'\\
\t e_3' & -\t e_3' & \t 0
\end{bmatrix}.
\]
As for the correction $\delta C=\delta_w C + \delta_x C$, it is composed of two terms, the first of which is due to the perturbation that induces the regular-distorted phase transition:
\[\label{eq:dwC}
\delta_w C = \begin{bmatrix}
\t 0 & -w_1\bar{\t e}_1'/a_1 & w_1\bar{\t e}_1'/a_1\\
w_2\bar{\t e}_2'/a_2 & \t 0 & -w_2\bar{\t e}_2'/a_2\\
-w_3\bar{\t e}_3'/a_3 & w_3\bar{\t e}_3'/a_3 & \t 0\\
\t 0 & -w_1\bar{\t e}_1'/b_1 & w_1\bar{\t e}_1'/b_1\\
w_2\bar{\t e}_2'/b_2 & \t 0 & -w_2\bar{\t e}_2'/b_2\\
-w_3\bar{\t e}_3'/b_3 & w_3\bar{\t e}_3'/b_3 & \t 0\\
\end{bmatrix},
\]
and the second of which is due to the fields being slowly varying in space:
\[\label{eq:dxC}
\delta_x C = \begin{bmatrix}
\t 0 & \t 0 & \t 0\\
\t 0 & \t 0 & \t 0\\
\t 0 & \t 0 & \t 0\\
\t 0 & \t e_1'\partial_1& \t 0\\
\t 0 & \t 0 & \t e_2'\partial_2\\
\t e_3'\partial_3 & \t 0 & \t 0
\end{bmatrix}.
\]
Last, the entries of the second-order correction $\delta^2 C$ will not be calculated as they turn out to be of no use for our purposes.

Similarly, displacements can be Taylor-expanded:
\[
\Phi = \Phi_0 + \delta\Phi + \delta^2\Phi + \dots,\quad \Phi=
\]
where $\Phi_0$ gathers the leading-order displacements, $\delta\Phi$ their first-order corrections and so on. As for the motion equation, it reads
\[\label{eq:ME}
-C'KC\Phi = -\omega^2M\Phi,
\]
where $K=\diag{\alpha_1,\alpha_2,\alpha_3,\beta_1,\beta_2,\beta_3}$ and $M=\diag{m_1\t I,m_2\t I,m_3\t I}$ are the diagonal rigidity and mass matrices and where $C'$ is the adjoint operator of $C$ obtained by transposing $C$ and mapping $\partial_j$ to $-\partial_j$. In the following, we derive an equation that governs the leading-order displacements $\Phi_0$ thus interpreted as the macroscopic motion equation. But first, the motion equation must be solved to leading and first orders.
\subsection{Leading and first order auxiliary problems}
Keeping only leading-order terms in the motion equation~\refeq{eq:ME} yields
\[
-C_0'KC_0 \Phi_0 = 0.
\]
Therefore, $\Phi_0'C_0'KC_0 \Phi_0 = 0$ and, by definiteness of $K$, $C_0\Phi_0=0$. We have seen in Subsection~\ref{subsec:FZMZ} that the solutions to this equation are periodic zero modes so that there exists a vector $\t U$ and an angle $\phi$ such that
\[
\Phi_0 = D\t U + T\phi.
\]

Then, keeping only the first-order terms entails
\[\label{eq:1storder}
-C_0'KC_0 \delta \Phi + \Psi = 0, \quad
\Psi = - C_0'K(\delta_wC + \delta_x C)(D\t U + T\phi).
\]
Thus, $\delta\Phi$ appears as a solution to a forced motion equation. Matrix $C_0$ being singular, the above equation admits solutions if and only if $\Psi$ is balanced in the sense of being orthogonal to all zero modes:
\[\label{eq:ortho}
D'\Psi = 0,\quad T'\Psi = 0.
\]
Alternatively, $\Psi$ is balanced if and only if it belongs to the range of matrix $C'_0$, which in turn is identical to the range of matrix
\[
G = \begin{bmatrix}
G_1 & G_2 & G_3
\end{bmatrix},\quad
G_1 = 
\begin{bmatrix}
 \t 0 \\ -\t e_1 \\ \t e_1
 \end{bmatrix},\quad
 G_2 = 
\begin{bmatrix}
\t e_2 \\ \t 0 \\ -\t e_2
\end{bmatrix},\quad
G_3 = 
\begin{bmatrix}
-\t e_3 \\ \t e_3 \\ \t 0
\end{bmatrix},
\]
given that $C_0' = \begin{bmatrix} G & -G\end{bmatrix}$. That is, $\Psi$ is a balanced loading if and only if it reads
\[
\Psi = G \psi, \quad \psi = \begin{bmatrix}
\psi_1 \\ \psi_2 \\ \psi_3
\end{bmatrix},
\]
where the $\psi_j$ are the generalized coordinates of $\Psi$ along the $G_j$. In equation~\ref{eq:1storder}, $\Psi$ is indeed balanced and a straightforward calculation shows that
\[\label{eq:psi}
\psi =
\begin{bmatrix}
(\gamma\beta_1-\alpha_1)w_1 \\
(\gamma\beta_2-\alpha_2)w_2 \\
(\gamma\beta_3-\alpha_3)w_3
\end{bmatrix}\phi
+
\frac{1}{3}
\begin{bmatrix}
\beta_1h_1\partial_1\\
\beta_2h_2\partial_2\\
\beta_3h_3\partial_3
\end{bmatrix}\phi
+
\begin{bmatrix}
\beta_1\scalar{\t e_1,\partial_1}\\
\beta_2\scalar{\t e_2,\partial_2}\\
\beta_3\scalar{\t e_3,\partial_3}
\end{bmatrix}\t U,
\]
where $\gamma=a_j/b_j$ is the $j$-independent similarity ratio and $h_j = \scalar{\t e_j,a_{j-1}\bar{\t e}_{j-1}}$ is the height of node $j$ in the triangle whose vertices are nodes $1$, $2$ and $3$ previously called triangle $\Delta$.

Therefore, a solution $\delta\Phi$ exists and is given by
\[
\delta\Phi = \Gamma\psi, \quad \Gamma = \begin{bmatrix}
\Gamma_1 & \Gamma_2 & \Gamma_3
\end{bmatrix},
\]
where $\Gamma_j$ is a solution to
\[
-C_0'KC_0\Gamma_j + G_j = 0.
\]
The $\Gamma_j$ are straightforward to determine from the above equation, first by solving for $KC_0\Gamma_j$, then for $C_0\Gamma_j$ and finally for $\Gamma_j$. Also, note that it is enough to calculate $\Gamma_1$ since $\Gamma_2$ and $\Gamma_3$ can be deduced by permutation symmetry. Skipping these steps, it comes that
\[\label{eq:SPFS}
\Gamma = -\frac{1}{2}
\begin{bmatrix}
\t 0 & \frac{a_3/h_2}{\alpha_2+\beta_2}\bar{\t e}_3 & \frac{a_2/h_3}{\alpha_3+\beta_3}\bar{\t e}_2 \\
\frac{a_3/h_1}{\alpha_1+\beta_1}\bar{\t e}_3 & \t 0 &  \frac{a_1/h_3}{\alpha_3+\beta_3}\bar{\t e}_1 \\
\frac{a_2/h_1}{\alpha_1+\beta_1}\bar{\t e}_2 &  \frac{a_1/h_2}{\alpha_2+\beta_2}\bar{\t e}_1 & \t 0
\end{bmatrix}.
\]

It is worth mentioning that the determined solution $\delta\Phi$ is not unique and can be modified by addition of an arbitrary periodic zero mode $D\delta\t U + T\delta\phi$. However, this will have no influence on what follows.
\subsection{Macroscopic equations of motion}
Keeping the second-order terms in the motion equation yields
\[\label{eq:delta2ME}
\begin{aligned}
-C_0' & KC_0 \delta^2 \Phi - C_0'K(\delta_wC + \delta_x C)\delta\Phi - (\delta_wC + \delta_x C)'KC_0\delta\Phi\\
& -C_0'K\delta^2C\Phi_0-(\delta_wC + \delta_x C)'K(\delta_wC + \delta_x C)\Phi_0 + F = -\omega^2M\Phi_0.
\end{aligned}
\]
Thus, $\delta^2\Phi$, just like $\delta\Phi$ before, is a solution to a forced motion equation and exists if and only if the orthogonality conditions~\refeq{eq:ortho} are enforced. The first one reads
\[\label{eq:m1}
- D'(\delta_x C)'KC_0\delta\Phi-D'(\delta_x C)'K (\delta_wC + \delta_x C)\Phi_0 + D'F = -\omega^2D'M\Phi_0.
\]
The second one is
\[\label{eq:m2}
\begin{aligned}
- T'(\delta_wC + \delta_x C)'KC_0\delta\Phi-T'(\delta_wC + \delta_x C)'K & (\delta_wC + \delta_x C)\Phi_0 \\
& + T'F = -\omega^2T'M\Phi_0.
\end{aligned}
\]
Both equations involve the leading-order displacements spanned by $\t U$ and $\phi$ and the applied body forces $F$, exclusively. Accordingly, they can be interpreted as a pair of macroscopic motion equations governing the macroscopic DOFs $\t U$ and $\phi$. Next, we write these equations in a form more suitable for interpretation, extract appropriate measures of strain and stress and reveal the constitutive law that relates them.
\subsection{Microtwist continuum}
The quantities involved in~\refeq{eq:m1} and~\refeq{eq:m2} can be fully evaluated simply by injecting therein the derived expressions \refeq{eq:D}, \refeq{eq:T}, \refeq{eq:comp0}, \refeq{eq:dwC}, \refeq{eq:dxC}, \refeq{eq:psi} and \refeq{eq:SPFS}. As a result, the macroscopic motion equations can be recast into the form
\[\label{eq:MME}
\begin{split}
-\omega^2(\rho\t U+\rho\bar{\t d}\phi)&= \t F +
\gt\nabla\cdot\left(\t C:\gt\nabla^s\t U + \t B\cdot\gt\nabla\phi + \t M\phi\right),\\
-\omega^2(\rho\scalar{\bar{\t d},\t U}+\eta\phi) &= \tau +
\gt\nabla\cdot\left(
\t B:\gt\nabla^s\t U + \t D\cdot\gt\nabla\phi + \t A\phi
\right)\\
&-\t M:\gt\nabla^s\t U -\t A\cdot\gt\nabla\phi - L\phi,
\end{split}
\]
where $\gt\nabla^s\t U$ is the symmetric part of the macroscopic displacement gradient, $\gt\nabla\phi$ is the twisting gradient, the operators $\cdot$ and $:$ symbolize simple and double contraction of tensors and $\gt\nabla\cdot$ is the divergence operator.

Vector $\t d=\sum_jm_j\t x_j/\sum_jm_j$ is the position vector of the center of mass of triangle $\Delta$ with respect to its geometric center and $\rho$ and $\eta$ are mass density and moment of inertia density
\[
\rho = \frac{\gamma^2}{ah(1+\gamma)^2}\sum_jm_j,\quad
\eta = \frac{\gamma^2}{ah(1+\gamma)^2}\sum_jm_j\norme{\t x_j}^2,
\]
where $ah/2\equiv a_jh_j/2$ is the area of triangle $\Delta$ and $ah(1+\gamma)^2/\gamma^2$ is the area of a unit cell and both are independent of $j$.

The vector-scalar pair $(\t F,\tau)$ is the resultant force-torque acting on a unit cell per unit cell area with respect to the geometric center of triangle $\Delta$. Its components read
\[
\t F = \frac{\gamma^2}{ah(1+\gamma)^2}\sum_j\t f_j,\quad
\tau =
\frac{\gamma^2}{ah(1+\gamma)^2}\sum_j\scalar{\bar{\t x}_j,\t f_j}.
\]

The involved effective tensors are given by
\[
\begin{alignedat}{3}
\t C &= \sum_j\frac{a_j}{h_j}k_j\t e_{jjjj},
\quad
&\t B &= \frac13\sum_ja_jk_j\t e_{jjj},\quad
&\t M &= \gamma\sum_j\frac{w_j}{h_j}k_j\t e_{jj},\quad\\
\t D &= \frac{ah}{9}\sum_jk_j\t e_{jj},\quad
&\t A &= \frac{\gamma}{3}\sum_j w_jk_j\t e_j,\quad
&L &= \frac{\gamma^2}{ah}\sum_jw_j^2k_j,
\end{alignedat}
\]
where $\t e_{jjjj}$, $\t e_{jjj}$ and $\t e_{jj}$ are the fourth, third and second tensorial powers of $\t e_j$ respectively, and $k_j = \alpha_j \beta_j/(\alpha_j + \beta_j)$. Accordingly, the above effective tensors are completely symmetric tensors of order four ($\t C$), three ($\t B$), two ($\t M$, $\t D$), one ($\t A$) and zero ($L$).

Alternatively, the macroscopic motion equations can be written as the balance equations
\[
-\omega^2(\rho\t U+\rho\bar{\t d}\phi)= \t F + \gt\nabla\cdot\gt\sigma,\quad
-\omega^2(\rho\scalar{\bar{\t d},\t U}+\eta\phi) =\tau + \gt\nabla\cdot\gt\xi + s,
\]
where $\gt\sigma$, $\gt\xi$ and $s$ are second, first and zero-order tensorial stress measures related to the strain measures $\gt\nabla^s\t U$, $\gt\nabla\phi$ and $\phi$ through the macroscopic constitutive law
\[\label{eq:tensors}
\begin{bmatrix}
\gt\sigma \\ \gt\xi \\ -s
\end{bmatrix}
=
\begin{bmatrix}
\t C & \t B & \t M \\
\t B & \t D & \t A \\
\t M & \t A & L
\end{bmatrix}
\begin{bmatrix}
\gt\nabla^s\t U \\ \gt\nabla\phi \\ \phi
\end{bmatrix},
\]

With the help of the divergence theorem, the motion equations can further be integrated over any domain $D$ with boundary $\partial D$ and outward unit normal $\t N$ to yield Euler's laws
\[
\begin{split}
-\omega^2\int_D\left(\rho\t U+\rho\bar{\t d}\phi\right)&=
\int_D \t F + \int_{\partial D} \gt\sigma\cdot\t N, \\
-\omega^2\int_D\left(\rho\scalar{\bar{\t d},\t U}+\eta\phi\right)&=
\int_D\tau + \int_{\partial D}\scalar{\gt\xi,\t N} + \int_D s.
\end{split}
\]
Knowing that $(\t F,\tau)$ is the resultant force-torque, the above equations readily provide an interpretation of the stress measures: $\gt\sigma$ is \emph{Cauchy's stress} whereby $\gt\sigma\cdot\t N$ yields the stress vector applied to a length element of normal $\t N$; $\gt\xi$ is \emph{couple stress} whereby $\scalar{\gt\xi, \t N}$ yields the torque per unit length applied to a length element of normal $\t N$; and $s$ is a \emph{hyperstress} counteracting the external body torque $\tau$.

We thus complete the description of the behavior of a general regular or weakly-distorted Kagome lattice, in the homogenization limit, as an enriched continuum with an extra DOF and additional measures of strain, stress and inertia. This enriched continuum is baptized \emph{the microtwist continuum}.
\subsection{Example: equilateral lattices}
\begin{figure}[ht]
\centering
\includegraphics[keepaspectratio,width=\textwidth]{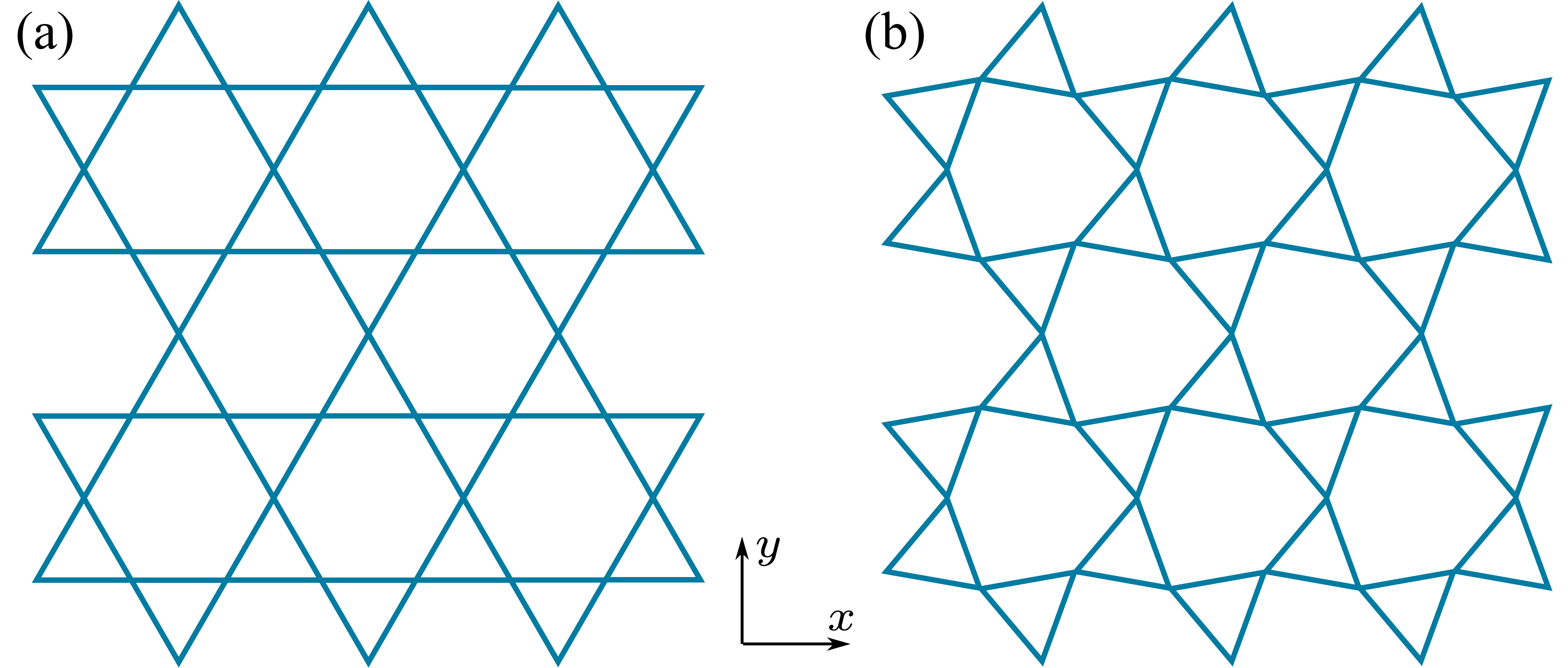}
\caption{Two equilateral Kagome lattices: (a) regular, (b) twisted.}
\label{fig:fig6}
\end{figure}

We readily exemplify the equations derived above in the case of Kagome lattices whose all edges are equal in length. We call such lattices \emph{equilateral}; see Figure~\ref{fig:fig6}. In particular, equilateral lattices are invariant by rotations of order $3$, have $j$-independent parameters and $\gamma=1$. Therefore, the effective tensors $\t C$, $\t D$ and $\t M$ are isotropic. Specifically, they are given by
\[
\t C:\gt\nabla^s\t U = \mu(2\gt\nabla^s\t U + \tr(\gt\nabla^s\t U) \t I),\quad
\t D = \frac{a^2}{3}\mu\t I,\quad
\t M = 4\frac{w}{a}\mu\t I,
\]
with $\mu=\sqrt{3}k/4$ and $k = k_1 = k_2 = k_3$. Above, we further assumed spring constants to be equal. In addition, $L = 8w^2\mu/a^2$ and vector $\t A$ vanishes. Assuming masses to be equal as well, the inertial coupling $\bar{\t d}$ vanishes whereas mass and moment of inertia densities simplify into
\[
\rho = \frac{\sqrt{3}m}{2a^2},\quad \eta = \frac{m}{2\sqrt{3}}.
\]
Last, the third-order effective tensor $\t B$, in the basis $(\t e_x\equiv \t e_1, \t e_y \equiv \bar{\t e}_1)$, has the components
\[
B_{xxx}=-B_{xyy}=-B_{yxy}=-B_{yyx}=\frac{a}{\sqrt{3}}\mu,\quad
B_{yxx}=B_{xyx}=B_{xxy}=B_{yyy}=0.
\]
These results are in agreement with the strain energy density postulated by \cite{Sun2012}.

The regular equilateral Kagome lattice is centrosymmetric. Nonetheless, its macroscopic constitutive law is not centrosymmetric since $\t B$, a third-order tensor, is non-zero. As a matter of fact, the action of inversion symmetry on the unit cell is trivial: it maps node $j$ onto itself. Thus, it has no influence on the macroscopic constitutive law. In contrast, reflection symmetry, parallel to $\t e_x$, permutes nodes $2$ and $3$ and has therefore a non-trivial action. The effect of reflection symmetry is visible on $\t B$: coefficients with an odd number of $y$ indices are all zero.

In any case, with these expressions, the motion equations can be fully expanded into
\[\label{eq:MER}
\begin{split}
-\omega^2\frac{\rho}{\mu} U_x &=
3U_{x,xx}+2U_{y,xy}+U_{x,yy} + \frac{a}{\sqrt 3}(\phi_{,xx}-\phi_{,yy}) + 4\frac{w}{a}\phi_{,x},\\
-\omega^2\frac{\rho}{\mu}U_y &=
3U_{y,yy}+2U_{x,xy}+U_{y,xx} - \frac{2a}{\sqrt 3}\phi_{,xy} + 4\frac{w}{a}\phi_{,y},\\
-\omega^2\frac{\eta}{\mu}\phi &=
\frac{a}{\sqrt 3}(U_{x,xx}-2U_{y,xy}-U_{x,yy}) + \frac{a^2}{3}(\phi_{,xx}+\phi_{,yy})
\\&-4\frac{w}{a}(U_{x,x}+U_{y,y})-8\frac{w^2}{a^2}\phi,
\end{split}
\]
and can be solved by prescribing appropriate boundary conditions where either $\t U$, $\phi$, $\gt\sigma\cdot\t N$ or $\scalar{\gt\xi,\t N}$, or a combination thereof is given, using the finite element method for instance; see Section \ref{sec:SPMC}.
\subsection{Discussion}\label{subsec:Discussion}
In conclusion of this section, several points are worth stressing. We do so in the following somewhat lengthy discussion.
\begin{enumerate}
\item
As it has more DOFs than dimensions, the microtwist medium qualifies as an enriched continuum in the sense of generalized continua \citep{Mindlin1964, CemalEringen1999}. The microtwist medium can be understood as a particular Cosserat medium where the microrotation DOF $\phi^\text{mr}$ and infinitesimal rotation $\gt\nabla\times\t U/2$ only appear in the combination $\phi=\phi^\text{mr}-\gt\nabla\times\t U/2$. Such a Cosserat medium would be unusual however as it would involve the second gradient of $\t U$, specifically $\gt\nabla(\gt\nabla\times\t U)$, through $\gt\nabla\phi$. This brings unnecessary formal complications; it seems then that Kagome lattices are more naturally understood as their own microtwist media. Microtwist media are also isomorphic to a subclass of Eringen's micromorphic media where microdeformation is restricted to a one dimensional space. Some refer to such a medium as a microdilatation medium; see, e.g., \cite{Forest2006}.
\item
\citet{Hutchinson2006} developed a static homogenization theory for regular equilateral Kagome lattices and other periodic trusses based on a kinematic hypothesis known as the Cauchy-Born hypothesis. It states that the displacements are the sum of one linear and one periodic field
\[
\t u^{l,m,n}_j = \t E \cdot \t x_j^{l,m,n} + \t u_j,
\]
the linear part being the result of an imposed uniform macroscopic deformation~$\t E$. In doing so, they precluded the presence of twisting gradients and neglected the coupling trio $\t B$, $\t D$ and $\t A$. Our model reduces to theirs when this approximation is implemented.

As a matter of fact, when the strain and stress fields are uniform, the twisting gradient $\gt\nabla\phi$ is necessarily null and the static equilibrium simplifies into $s=0$. Solving for $\phi$ entails $\phi = -\t M:\gt\nabla^s\t U/L$. Finally, the reduced stress-strain relationship reads
\[
\gt\sigma = \t C^* :\gt\nabla^s\t U;\quad
\t C^* = \t C - \frac{1}{L}\t M\tp\t M.
\]
This generalizes the results of \citet{Hutchinson2006} to arbitrary regular and weakly-distorted Kagome lattices. Using the elasticity tensor $\t C^*$ is appealing as it greatly simplifies the constitutive law. Nonetheless, recall that its use is only justified for static uniform fields. Otherwise, the trio $(\t B,\t D,\t A)$ cannot be justifiably neglected. Taking this coupling into account will in fact significantly improve the quality of the predictions of the effective medium theory; various quantitative demonstrations are suggested in the following section.
\item
In the preceding derivations, nodes were assumed to behave like perfect hinges. The consequence is that variations of angles between the bonds meeting at a given node cost no elastic energy at all. It could be of interest however to inspect the mechanics of Kagome lattices with elastic hinges as they are expected to be better models of real structures; see Figure~\ref{fig:fig7}. Taking the influence of elasticity in the hinges turns out to be remarkably simple so long as the hinges are soft. Indeed, in that case, it is enough to change the expression of the effective parameter $L$ into
\[
L = \kappa + \frac{\gamma^2}{ah}\sum_jw_j^2k_j
\]
where $\kappa$ is an effective torsional spring constant function of geometry and of the elasticity moduli of the hinges. A proof is outlined in Appendix~C.

\begin{figure}[ht]
\centering
\includegraphics[keepaspectratio,width=0.7\textwidth]{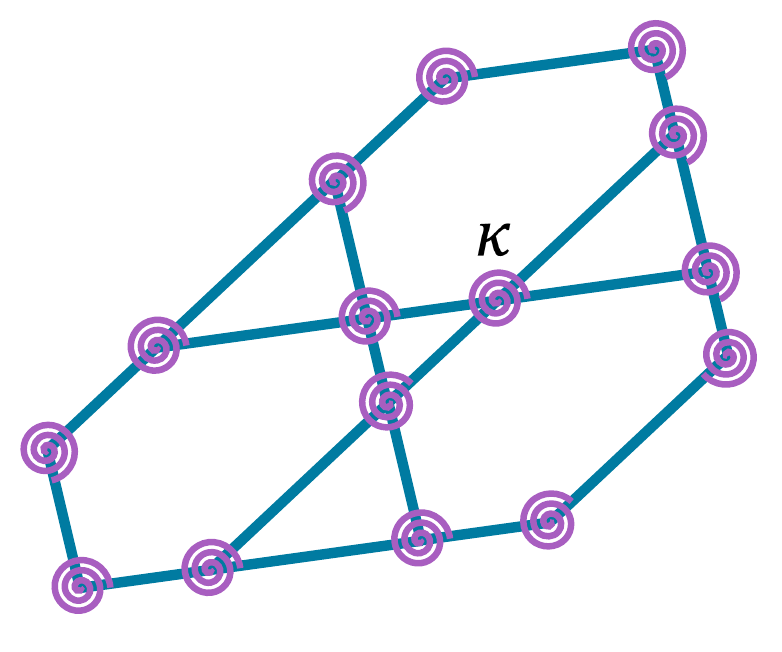}
\caption{A Kagome lattice with rotational springs $\kappa$ denoted by purple spirals.}
\label{fig:fig7}
\end{figure}
\item
The quadratic form of strain energy density $\epsilon$ is
\[
\epsilon = \frac{\gt\sigma:\gt\nabla^s\t U + \gt\xi\cdot\gt\nabla\phi - s\phi}{2},
\]
where stresses are linear combinations of strains following the constitutive law of the microtwist continuum. Skipping calculations, its expression can be recast into
\[\label{eq:strE}
\epsilon = \sum_j \frac{k_j}{2ah}\left(
a_j\t m_{jj}:\gt\nabla^s \t U + 
\frac{ah}{3}\t m_j\cdot\gt\nabla\phi +
\gamma w_j \phi
\right)^2 + \frac{\kappa}{2}\phi^2
\]
where it is clear that it is non-negative. Definiteness however completely relies on the elastic constants $k_j$ and $\kappa$ being non-null. In particular, when the hinges are perfect ($\kappa=0$), strain energy is semi-definite and therefore allows for microstructural zero modes to manifest on the macroscopic scale.
\item
P-symmetry stipulates that over a centrosymmetric domain, there corresponds to each solution $(\t U(\t x),\phi(\t x))$ a space-inverted solution $(\t U(-\t x), \phi(-\t x))$. P-symmetry therefore requires that strain energy be invariant under the formal substitution $\gt \nabla\mapsto -\gt\nabla$. Then, it is easily seen from equation~\refeq{eq:strE} that a Kagome lattice is P-symmetric, in the homogenization limit, if and only if all $w_j$ vanish, i.e., if and only if it is regular. Conversely, all distorted lattices are P-asymmetric and therefore polarized.

It is of interest to note how strain energy is always invariant under $(\gt \nabla,w_j)\mapsto (-\gt\nabla,-w_j)$. This is similar to how in certain physical systems, P-symmetry breaks but CP-symmetry survives where ``C'' stands for charge conjugation. This allows us to infer the effective constitutive tensors responsible for P-asymmetry: they are odd functions of the $w_j$. We conclude that the tensors $\t A$ and $\t M$ are at the origin of polarization effects in the microtwist medium.

Moreover, tensor $\t M$ is alone responsible for polarization effects in the bulk. Indeed, tensor $\t A$ only appears in the combination $\gt\nabla\cdot(\t A \phi) - \t A\cdot\gt\nabla\phi$ which vanishes except when the lattice is modulated in space, i.e., for $\t A=\t A(\t x)$. In any case however, both $\t M$ and $\t A$ contribute to polarization near interfaces or edges since they would be involved in writing the corresponding continuity and boundary conditions weighing on $\gt\sigma$ and $\gt\xi$.

\end{enumerate}
%
\section{Performance of the microtwist medium}\label{sec:SPMC}
Having derived the equations of the microtwist continuum, it is time to inquire whether it is faithful in its predictions of the elastic behavior of different Kagome lattices, be them regular or weakly-distorted. Here, we assess the performance of the microtwist model in the contexts of dispersion diagrams, zero modes and polarization effects.
\subsection{Prediction of dispersion relations}\label{sub:BLZM}
Free harmonic plane waves propagated through the Kagome lattice exist at specific frequencies $\omega$ and wavenumbers $\t q$ solution to the dispersion relation
\[\label{eq:discDisp}
\det\left( C(\t q)'KC(\t q)-\omega^2 M \right) = 0.
\]
The Kagome lattice having six DOFs per unit cell, there exists six solution frequencies $\omega=\omega_d(\t q)$, $d=1,2,\dots 6$, for any given wavenumber $\t q$. The microtwist continuum, having three DOFs per unit cell, will be able, at best, to account for the lowest three of them $\omega=\tilde{\omega}_d(\t q)$, $d=1,2,3$. These frequencies are obtained by injecting
\[\label{eq:PWE}
\t U(\t x,t) = \t U \exp(i\scalar{\t q,\t x}-i\omega t),\quad
\phi(\t x,t) = \phi \exp(i\scalar{\t q,\t x}-i\omega t),
\]
in equation~\refeq{eq:MME} under zero loading and solving the resulting dispersion relation
\[\label{eq:MTdisp}
\det\left(
\begin{bmatrix}
\t q\cdot\t C\cdot\t q & \t q\cdot\t B\cdot \t q - i\t q\cdot\t M\\
(\t q\cdot\t B\cdot \t q - i\t q\cdot\t M)' & \t q\cdot\t D\cdot\t q + L
\end{bmatrix}
-\omega^2\rho
\begin{bmatrix}
\t I & \bar{\t d} \\ \bar{\t d}' & \eta/\rho
\end{bmatrix}
\right)=0.
\]
Hereafter we draw a comparison between the two, discrete and microtwist, models. For reference, we also include the dispersion diagrams of the effective Cauchy continuum.
\begin{figure}[ht!]
\centering
\includegraphics[keepaspectratio,width=\textwidth]{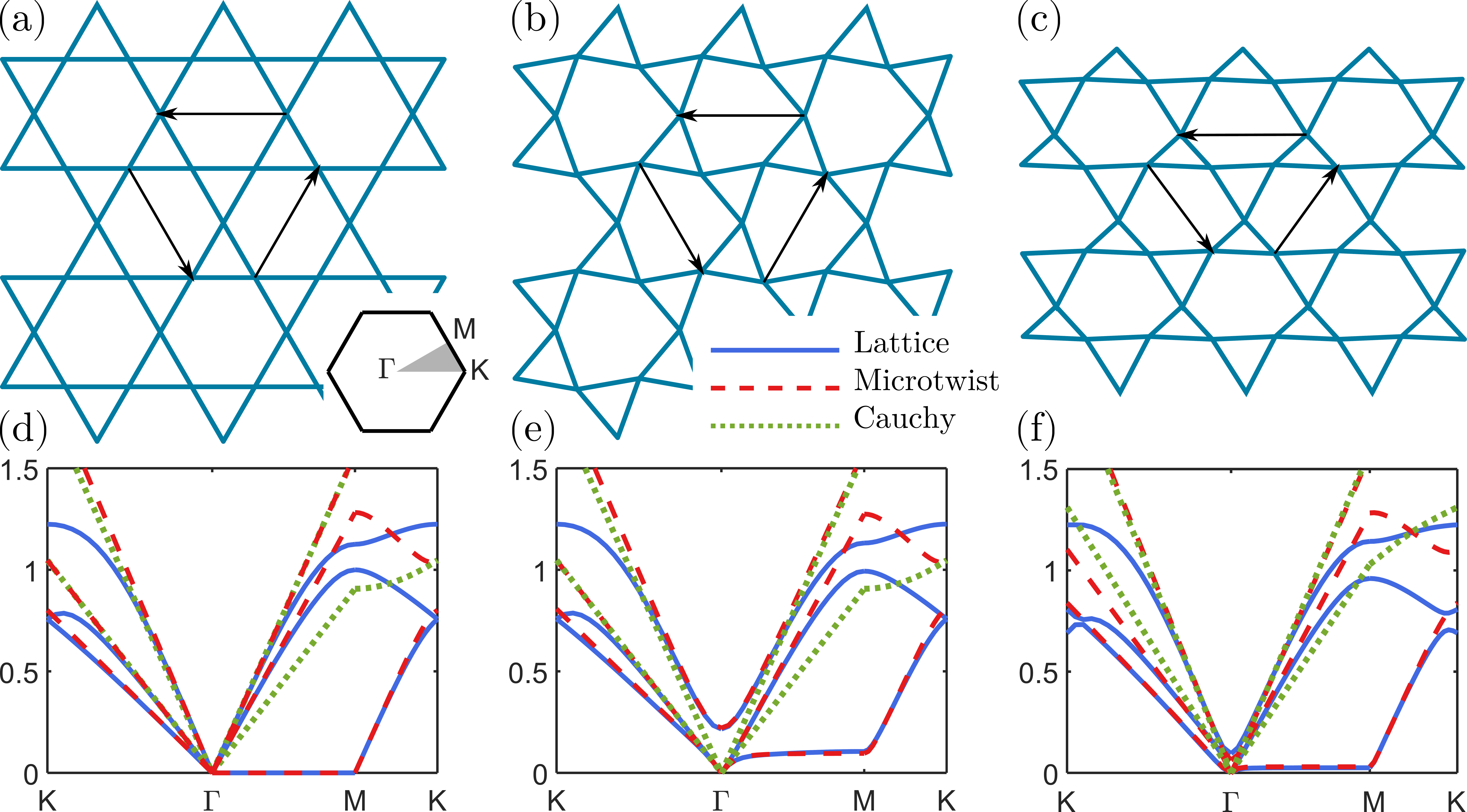}
\caption{Kagome lattices and their normalized dispersion diagrams: (a, d) regular; (b, e) twisted; and (c, f) topological. The used numerical values are: $a_j=b_j=1$ and $\alpha_j=\beta_j=1$; $\theta_j=0^\circ$ for (a); $\theta_j=-5^\circ$ for (b); and $(\theta_1,\theta_2,\theta_3)\approx(-2^\circ,1^\circ,-5^\circ)$ for (c).}
\label{fig:disp}
\end{figure}

\begin{figure}[ht!]
\centering
\includegraphics[keepaspectratio,width=\textwidth]{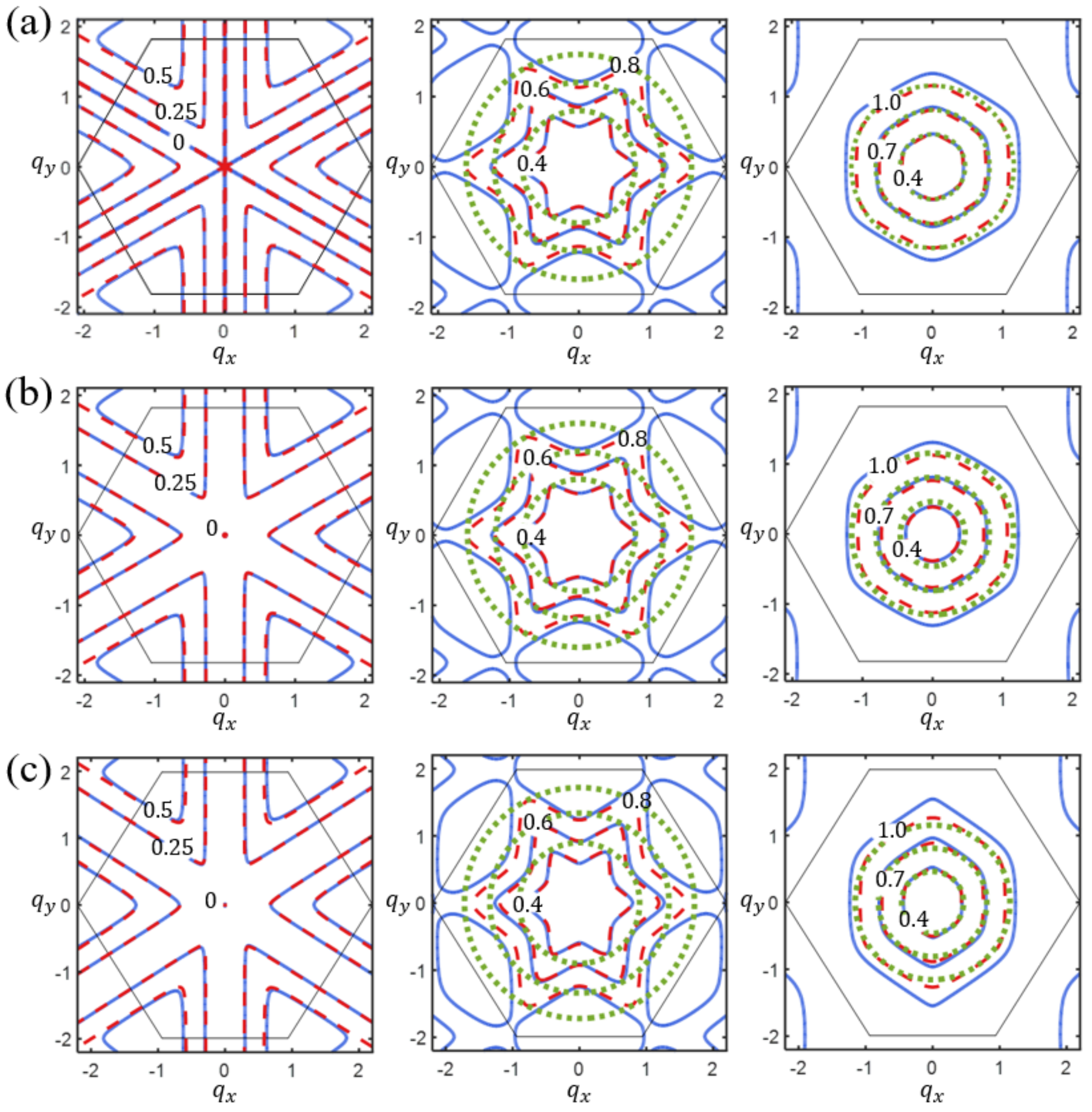}
\caption{Isofrequency contours of the first (left), second (middle) and the third (right) dispersion surfaces using the lattice, microtwist and Cauchy models for the regular (a), twisted (b) and topological (c) lattices of Figure~\ref{fig:disp}. The traced hexagons delimit the respective first Brillouin zones.}
\label{fig:iso}
\end{figure}

Thus let us consider the three Kagome lattices (a), (b) and (c) of Figure~\ref{fig:disp}. While lattice (a) is regular, lattices (b) and (c) are weakly-distorted. In comparison, the distortion parameters of lattice (b) are all of the same sign whereas for lattice (c) some are positive and others are negative. In what follows, we refer to lattice (a) as regular, to (b) as twisted, and to (c) as ``topological'', for reasons that shall become clear. In any case, across all three lattices, the dispersion diagrams (d-f) show satisfactory agreement between the discrete and microtwist models up to frequencies comparable to the cutoff frequencies of the three lowest dispersion branches and that for small to medium wavenumbers. By contrast, the Cauchy model systematically misses one dispersion branch, the one corresponding to the twisting motion, and in some directions, e.g., $\Gamma$M direction, misses the shear acoustic branch. These observations hold for the isofrequency contours shown on Figure~\ref{fig:iso}. Note again how the Cauchy model completely omits the first dispersion surface (left); in particular, it exhibits no traces of the zero modes of the regular lattice (a). The Cauchy model does well in particular highly symmetric directions over the second dispersion surface (middle) and is satisfactory overall for the third surface (right). However, the Cauchy model fails to describe directional behavior of the second dispersion surface. By contrast, the microtwist model appears consistently accurate across all three surfaces for wavelengths as small as two unit cells.
\subsection{Continuum characterization of zero modes}\label{subsec:zeromodes}
Zero modes, over a continuum, can be reasonably defined as configurations producing zero stress measures:
\[
\gt\sigma = \t 0,\quad \gt\xi = \t 0, \quad s = 0.
\]
Equivalently, zero modes produce zero strain energy: $\epsilon=0$. Thus, in light of expression~\refeq{eq:strE}, zero modes only exist in the absence of elasticity in the hinges (i.e., $\kappa=0$) and, in that case, are solutions to
\[\label{eq:freeenergy}
a_j\t e_{jj}:\gt\nabla^s \t U + 
\frac{ah}{3}\t e_j\cdot\gt\nabla\phi +
\gamma w_j \phi
=
0, \quad j=1,2,3.
\]
The above system of linear partial differential equations provides a continuum characterization of the zero modes of regular and weakly-distorted Kagome lattices. It is noteworthy that this characterization is independent of the elastic moduli $k_j$, zero modes being representative of configurations that do not stretch any bonds.

\begin{figure}[ht!]
\centering
\includegraphics[keepaspectratio,width=\textwidth]{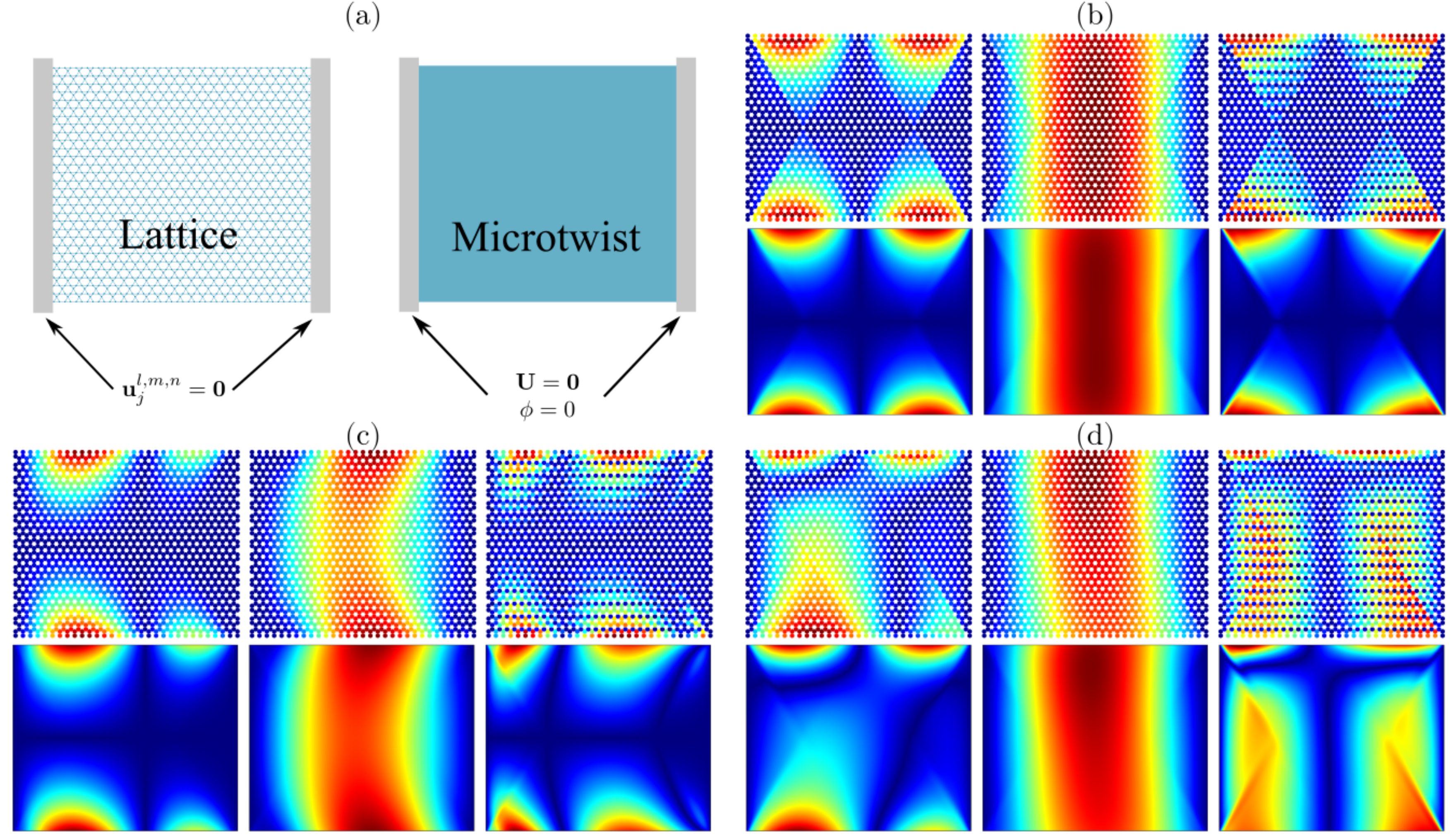}
\caption{Comparison of fundamental near-zero mode prediction from lattice model and microtwist model: (a) imposed boundary conditions; (b-d) mode shape for the discrete (top) and the continuum (bottom) models resolved into the three components $U_x$ (left), $U_y$ (middle) and $\phi$ (right).}
\label{fig:zeroModes}
\end{figure}

It is possible to numerically solve the above system under appropriate boundary conditions. Alternatively, it is more convenient to obtain approximate zero modes by minimizing strain energy in the presence of a small residual elastic energy stored in the hinges (i.e., for $0<\kappa\sim 0$). Thus, we consider a rectangular sample freely vibrating under Dirichlet left and right boundary conditions and free top and bottom boundaries as shown on Figure~\ref{fig:zeroModes}a. We then calculate the eigenmode of lowest energy for both the discrete and continuum models. This fundamental eigenmode becomes a zero mode in the limit $\kappa\to 0$. Three components corresponding to $U_x$, $U_y$ and $\phi$ are extracted from the eigenmode's shape and are plotted as normalized color maps on Figure~\ref{fig:zeroModes}b-d for the same regular, twisted and topological lattices as before. In all of these cases, the microtwist continuum predicts well the mode shape of the approximate zero mode.

It is of interest here to recall how distortions influence the shape of zero modes. As illustrated earlier on Figure~\ref{fig:fig3}, and again on Figures~\ref{fig:disp} and~\ref{fig:iso}, distorted lattices such that all $\theta_j\neq 0$ admit no Floquet-Bloch zero modes, precluding rigid body motions. Accordingly, large-enough samples of such lattices admit no bulk zero modes. For instance, the zero mode calculated for the regular lattice on Figure~\ref{fig:zeroModes}b spans the bulk of the sample but shifts towards the top and bottom edges following the introduction of the distortion as seen on Figure~\ref{fig:zeroModes}c, d. A closer look reveals in fact that the twisted lattice redistributes the zero mode evenly between the top and bottom edges whereas the topological lattice prefers the top edge. This phenomenon of polarization appears to be ubiquitous in distorted Kagome lattices and is most pronounced in those lattices qualified as ``topological'' by \cite{Kane2014}. Polarization effects are explored hereafter in more detail.
\subsection{General polarization effects}
In order to understand how polarization effects emerge, we revisit the field equations~\refeq{eq:freeenergy} of the zero modes of the microtwist continuum written for a plane wave of wavenumber $\t q$, namely
\[
ia_j\scalar{\t e_j,\t q}\scalar{\t e_j,\t U} + 
\left(i\frac{ah}{3}\scalar{\t e_j,\t q}+
\gamma w_j\right)\phi
=
0, \quad j=1,2,3.
\]
Non-trivial solutions then exist when the zero-determinant condition
\[
\det\begin{bmatrix}
ia_1\scalar{\t e_1,\t q}\t e_1' & 
i\frac{ah}{3}\scalar{\t e_1,\t q}+
\gamma w_1 \\
ia_2\scalar{\t e_2,\t q}\t e_2' & 
i\frac{ah}{3}\scalar{\t e_2,\t q}+
\gamma w_2 \\
ia_3\scalar{\t e_3,\t q}\t e_3' & 
i\frac{ah}{3}\scalar{\t e_3,\t q}+
\gamma w_3
\end{bmatrix} = 0
\]
is met. Equivalently, $\t q$ is solution to the zero-frequency dispersion relation
\[\label{eq:dispZMM}
ah \prod_j \scalar{\t e_j,\t q}
- i \gamma\sum_j w_j\prod_{k\neq j} \scalar{\t e_k,\t q} = 0.
\]
Now real solutions $\t q$ have been characterized earlier; Figure~\ref{fig:fig3} summarizes their whereabouts. Hereafter, we focus on the case $\prod_j w_j\neq 0$ for which no real non-zero solutions exist. Thus, we look for complex non-real solutions in the particular form $\t q = q_x\t e_x + q_y\t e_y$ where $q_x$ is real and non-zero, $q_y = q_R + iq_I$ is complex with real part $q_R$ and imaginary part $q_I$ and where $\t e_y$ is inclined with respect to $\t r_1$ by an angle $\zeta$. Note that $q_I$ is necessarily non-zero since otherwise $\t q$ would be real and non-zero, which is impossible here.

Such complex solutions correspond to zero modes whose amplitude remains bounded in the $x$-direction and either decays or grows exponentially in the $y$-direction. Suppose that a Kagome lattice occupies the $y>0$ half-space, then we call ``admissible'' solutions that decay deep inside the lattice, that is solutions for which $q_I$ is positive. Equation~\refeq{eq:dispZMM} being of degree three, in general, any given free surface, of inclination $\zeta$, will admit up to three admissible zero modes per wavenumber $q_x$. A key observation here is that to any admissible solution $q_y=q_y(q_x,\zeta)$, there corresponds a non-admissible complex conjugate solution $q_y^c=q_y(q_x,\zeta+\pi)$. This is because $q_y$ and $q_y^c$ have opposite imaginary parts. Hence, in general, if a free surface of inclination $\zeta$ admits $Z$ zero modes, then the opposite free surface of inclination $\zeta+\pi$ admits $3-Z$ zero modes. But $Z$, being an integer, cannot equal $3-Z$. Therefore, two opposite free surfaces will in general admit different numbers of zero modes. This imbalance is a symptom of a broken P-symmetry and gives rise to polarization effects along almost all directions of distorted Kagome lattices.
\subsection{Macroscopic topological polarization}
\begin{figure}[ht!]
\centering
\includegraphics[keepaspectratio,width=\textwidth]{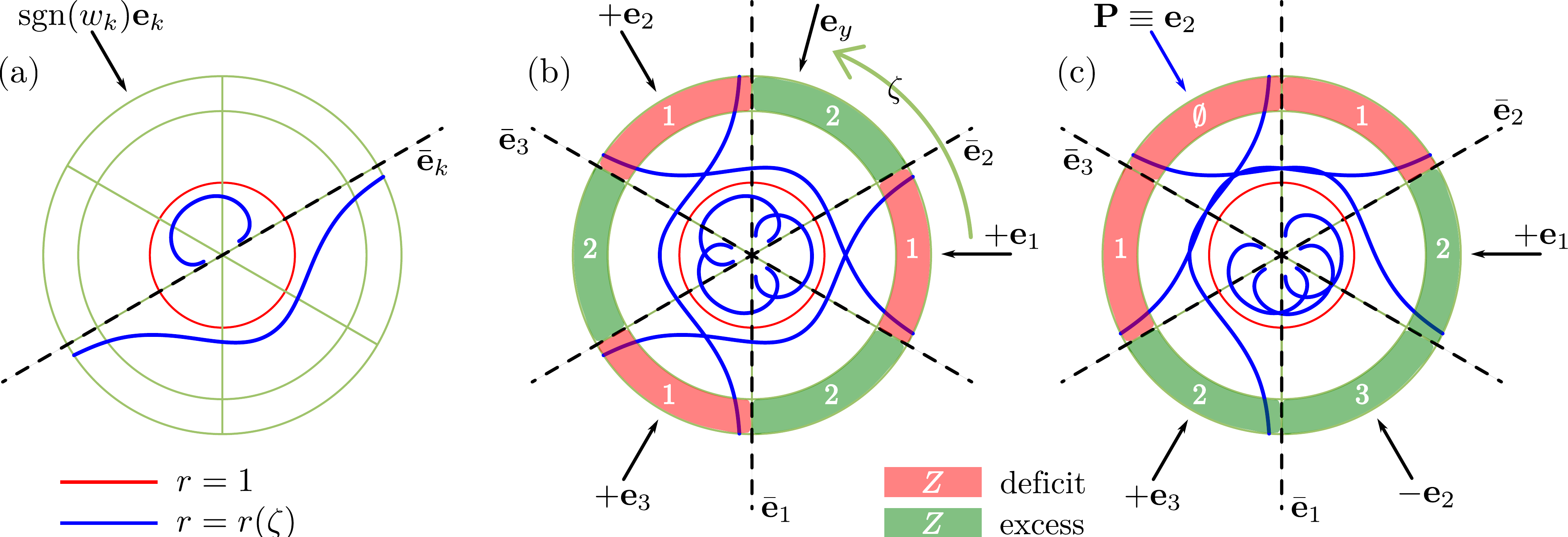}
\caption{Polar plots of the decay/amplification factor $r=r(\zeta)$: (a) generic plot of branch $k$; (b) generic plot for all branches in the trivial configuration $(w_j>0)$; (c) generic plot for all branches in the topological configuration $(w_1,w_3>0,w_2<0$). Sectors are color-coded to indicate when they exhibit an excess or a deficit of zero modes compared to the opposite sector; note how in the topological configuration, sectors with an excess of zero modes are adjacent and form a maximal region or normal $\t P$.}
\label{fig:polarization}
\end{figure}

Some lattices exhibit more pronounced polarization effects than others. To understand why, let us sweep angles between $\zeta$ and $\zeta+\pi$. As $Z$ changes to $3-Z$, we realize that some initially admissible zero modes become non-admissible, and conversely. Therefore, there must be angles at which some solutions $q_I$ change signs. Since $q_I$ cannot vanish, the only way for it to change signs is by going to infinity. To identify such wavenumbers, we look for particular solutions to equation~\refeq{eq:dispZMM} such that $q_I\to\pm\infty$. In that case, to leading order, $\t q \sim iq_I\t e_y$ implies
\[
-iah q_I^3\prod_j \scalar{\t e_j,\t e_y}
+ i q_I^2\gamma\sum_j w_j\prod_{k\neq j} \scalar{\t e_k,\t e_y} = 0,
\]
and
\[
q_I = \frac{\gamma}{ah}\sum_j \frac{w_j}{\scalar{\t e_j,\t e_y}}.
\]
The above solution is only accurate if consistent with the premise $q_I\to\pm\infty$, i.e., only in the vicinity of normal directions such that
\[
\scalar{\t e_j,\t e_y} = 0
\]
for some $j=k$. Accordingly, the solution further simplifies into
\[
q_I = \frac{\gamma}{ah}\frac{w_k}{\scalar{\t e_k,\t e_y}}.
\]
It is insightful to draw a generic polar plot $r=r(\zeta)$ of the decay factor $r(\zeta)= \exp\left(-q_I(q_x,\zeta)/q_x\right)$, where $q_x$, taken positive, is used as an arbitrary normalization factor; see Figure~\ref{fig:polarization}a. Indeed, we know that $r$ approaches $0$ as $q_I$ approaches $+\infty$, i.e., as $\t e_y$ approaches $\pm\bar{\t e}_k$ from the same side as $\sign(w_k)\t e_k$; between these two extremes, $r$ necessarily remains smaller than $1$ since $q_I$ cannot vanish. Conversely, $r$ approaches $+\infty$ when $q_I$ approaches $-\infty$, i.e., as $\t e_y$ approaches $\pm\bar{\t e}_k$ from the opposite side of $\sign(w_k)\t e_k$; between these two extremes, $r$ necessarily remains greater than $1$ since, again, $q_I$ cannot vanish.

There exists of course one such branch for each $k$. When the various branches for $k=1,2,3$ are combined, two qualitatively different configurations arise. In one, the vectors $\sign(w_k)\t e_k$ point into different half-planes; we call this configuration ``trivial'' (Figure~\ref{fig:polarization}b). In the other, all vectors $\sign(w_k)\t e_k$ point into the same half-plane; we call this configuration ``topological'' (Figure~\ref{fig:polarization}c). Alternatively, a configuration is trivial when all $w_k$ have the same sign and is topological otherwise, i.e., if one $w_k$ has a different sign from the other two $w_j$.

Either way, it is possible to identify six angular sectors delimited by the $\bar{\t e}_k$ and to gauge whether within each sector there is an excess or a deficit of admissible zero modes. Remarkably, in the topological configuration, sectors with excess of zero modes aggregate and form a complete half-plane. We define a macroscopic polarization vector as the out-going unit normal $\t P$ of said half-plane. Therefore, $\t P=-\sign(w_k)\t e_k$ if and only if $w_k$ has an opposite sign to the other two $w_j$; see, e.g., Figure~\ref{fig:polarization}c. By contrast, in the trivial configuration, sectors with excess of zero modes are disjoint and interspersed by sectors in deficit. In this case, i.e., when all $w_k$ have the same sign, we set $\t P=\t 0$.

Polarization $\t P$ defined in this manner is independent of $q_x$. Moreover, $\t P$ is independent of the particular geometric and constitutive parameters of the underlying lattice; it only depends on the signs of the perturbations $w_j$. Thus $\t P$ is an example of a ``topological invariant''. In order to change $\t P$, one family of zero modes must migrate from one free surface to its opposite meaning that the lattice must at some point allow for bulk zero modes to exist. But bulk zero modes have been excluded by the condition $\prod_j w_j\neq 0$. In other words, zero modes can only migrate between opposite free surfaces when a $w_j$ reaches zero and changes signs which is, again, excluded by the same condition.

It is understandable then that topologically polarized lattices, i.e. lattices with $\t P\neq\t 0$, exhibit more pronounced polarization effects than trivial lattices. Indeed, the existence of a non-zero $\t P$ implies an excess of zero modes along a maximal region, i.e., a half-plane.

Note that this macroscopic notion of topological polarization is only valid in the limit of small distortions $w_j$. For large enough $w_j$, zero modes localize over thin boundary layers and can no longer be captured using the present continuum theory, not in its current form at least. Based on a study of the discrete lattice, \cite{Kane2014} introduced and interpreted a topological polarization vector $\t P^{KL}$ which similarly serves to pinpoint free surfaces with an excess of zero modes. Their analysis led to the elegant formula
\[
\t P^{KL} = -\frac{1}{2}\sum_k \sign(w_k)\t r_k.
\]
Retrieving $\t P^{KL}$ on a macroscopic level was a principal motivation behind the present work. Indeed, it is straightforward to check that $\t P^{KL}$ and $\t P$ point in the same direction. That being said, it is important to stress that $\t P^{KL}$ and $\t P$ do not count the same zero modes. Specifically, the construction of $\t P$ is based on counting macroscopic zero modes, i.e., those zero modes which decay or grow slow enough across many unit cells so as to survive a micro-to-macro scale transition. Vector $\t P^{KL}$ on the other hand takes into account all zero modes however localized.
\subsection{Indentation tests}
Asymmetric distributions of zero modes are expected to cause a polarized elastic response. Here, we investigate the polarized elastic response of a topological lattice and assess whether the microtwist theory is capable of accurately reproducing that response on a macroscopic level.
\begin{figure}[ht!]
\centering
\includegraphics[keepaspectratio,width=\textwidth]{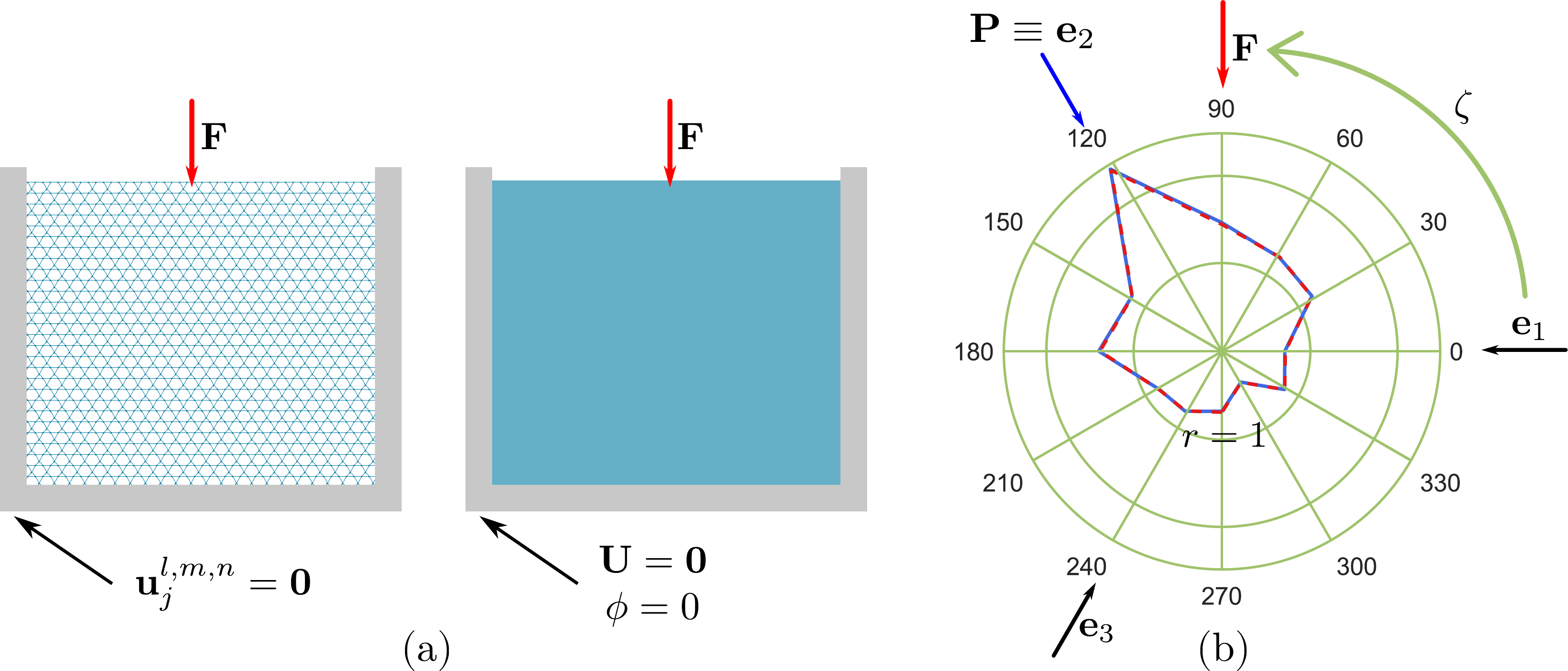}
\caption{Indentation tests simulated for free surfaces of different inclinations $\zeta$: (a) boundary conditions; (b) polar plot of the relative surface stiffness. Configuration (a) corresponds to $\zeta=\pi/2$ where the top edge is free and the force is downward.}
\label{fig:indentation}
\end{figure}

Consider a rectangular sample of a topological lattice with one free edge and three constrained edges. Like before, we let $\t e_y$ be the inward unit normal along the free edge and call $\zeta$ the angle it makes with $\t r_1$; vector $\t e_x$ is aligned with the free edge. For angles $\zeta$ sweeping the range $[0,2\pi)$, we apply an inward force $\t F$ at the midpoint of the free edge and calculate, using FEA, the displacement at the same point and in the same direction (Figure~\ref{fig:indentation}a).  The force to displacement ratio defines a surface stiffness $S(\zeta)$; the polar plot $r(\zeta) = S(\zeta)/S(\zeta+\pi)$ compares the stiffnesses of two opposite free surfaces and allows to gauge how polarized the elastic response is. A similar plot is made using the microtwist continuum. The two plots match satisfyingly (Figure~\ref{fig:indentation}b).

For this particular example, $\t P=\t e_2$ since $w_2$ is negative whereas $w_1$ and $w_3$ are positive. Thus, zero modes overpopulate edges where the in-going unit normal forms an obtuse angle with $\t P$. These correspond to angles $\zeta$ between $-150^\circ$ and $30^\circ$, approximately. For all of these angles, the relative surface stiffness is lower than 1 signaling a relative softening of these edges. By contrast, the remaining edges, deserted by zero modes, exhibit a relative hardening. The maximum is reached in the direction of $\t P$, approximately. It is worth noting here, that in order to guarantee the well-posedness of the indentation problem, simulations are performed in the presence of a small residual elasticity in the hinges $0<\kappa\sim 0$.
\section{Conclusion}
In this paper, we developed a theory of elasticity, called microtwist elasticity, that can capture the zero modes and topological polarization of Kagome lattices on a macroscoptic scale. Performance of the proposed theory is validated against the discrete model in a number of problems including determining the P-asymmetric distribution of zero modes, calculating the dispersion relations and quantitatively predicting the polarized indentation response of finite samples. The theory also permits to establish a hierarchy of isostatic Kagome lattices depending on how they are geometrically distorted.
\begin{itemize}
\item
Lattices with all distortions $w_j=0$ are regular and non-polarized; lattices with some $w_j\neq 0$ are distorted and polarized.
\item
Lattices with all $w_j\neq 0$ are further gapped at $\omega=0$, $\t q\neq\t 0$: if all $w_j$ are of the same sign then the lattice is trivially polarized; if the $w_j$ are not all of the same sign then the lattice is topologically polarized.
\end{itemize}

Microtwist elasticity is capable of distinguishing all of these notions and of producing quantitatively accurate predictions when the distortion parameters are kept small. The theory extends easily to nearly-isostatic Kagome lattices, i.e., with next-nearest-neighbor interactions or elasticity in the hinges, as long as the elastic constants of the bonds breaking isostaticity are kept small. It also extends to other isostatic lattices that are on the brink of a regular-distorted, or polarized-unpolarized, phase transition.

It is of interest to see if and how the theory could be extended to strongly distorted lattices in connection to the work of \cite{Sun2019}, of \cite{Saremi2020} and of \cite{Marigo2016,Marigo2017}. It is equally interesting to explore how polarization influences the dynamic behavior and to investigate the role of the coupling measure of microinertia $\bar{\t d}$. Finally, experimental efforts characterizing the polarized behavior of Kagome lattices are much needed.
\section*{Acknowledgments}
This work is supported by the NSF CMMI under Award No.\,1930873 with Program Manager Dr.\,Nakhiah Goulbourne, the Air Force Office of Scientific Research under Grant No.\,AF 9550-18-1-0342 with Program Manager Dr.\,Byung-Lip (Les) Lee and the Army Research Office under Grant No.\, W911NF-18-1-0031 with Program Manager Dr.\,Daniel P\,Cole.  

\section*{Appendix A. The rank of the compatibility matrix}\label{app:rank}
We have seen that, for periodic configurations in a general Kagome lattice, the compatibility matrix reads
\[
C =
\begin{bmatrix}
\t 0 & -\t m_1' & \t m_1'\\
\t m_2' & \t 0 & -\t m_2'\\
-\t m_3' & \t m_3' & \t 0\\
\t 0 & -\t n_1' & \t n_1'\\
\t n_2' & \t 0 & -\t n_2'\\
-\t n_3' & \t n_3' & \t 0
\end{bmatrix}.
\]
By the rank-nullity theorem, the number of zero modes is equal to $Z = 6-\rank C$ where $6$ is the dimension of $C$ and $\rank C$ its rank. On one hand, translations systematically provide two linearly independent periodic zero modes so that $Z\geq 2$. On the other hand, the first three lines of matrix $C$ are necessarily linearly independent because, for instance, $\t m_2$ and $\t m_3$ can never be parallel. Thus, $\rank C\geq 3$ leaving us with two possibilities: $(Z,\rank C)=(2,4)$ or $(3,3)$.

If there exists a $j$ such that $\t m_j$ and $\t n_j$ are misaligned, then $Z=2$. Indeed, say, for the sake of argument, that $j=1$. Then, the extracted $4\times 4$ matrix
\[
\begin{bmatrix}
-\t m_1' & \t m_1'\\
\t 0 & -\t m_2'\\
\t m_3' & \t 0\\
-\t n_1' & \t n_1'
\end{bmatrix}
\]
has a non-zero determinant since otherwise $\t m_2$ and $\t m_3$ would be aligned. Therefore, $C$ is of rank $4$ and $Z=2$. In contrast, if for all $j$, $\t m_j$ and $\t n_j$ are parallel, then $Z=3$. As a matter of fact, $\t m_j$ and $\t n_j$ being aligned and unitary means they are equal and opposite. In that case, the last three lines of $C$ are exactly the opposites of the first three ones and the rank of $C$ cannot exceed its lower bound of $3$ nor can $Z$ decrease below its upper bound of $3$.
\section*{Appendix B. Existence of Floquet-Bloch zero modes}\label{app:existence}
In the main text, it was proven that a Floquet-Bloch zero mode of wavenumber $\t q$ such that $P_j=1$ exists if and only if $\theta_j=0$. But then it is unclear whether other solutions exist where $P_j\neq 1$ for all $j$. Here, we provide a sufficient condition under which there exist no solutions to~\refeq{eq:PZM} outside of the lines $P_j=1$.

Thus, let $\t q$ be a real wavenumber such that $P_j\neq 1$ for all $j$. Divide~\refeq{eq:PZM} by $\prod_j(1-P_j)$ yielding
\[\label{eq:pre}
\sum_j \scalar{b_j\t n_j,\bar{\t x}_{j+1}} +
\sum_j \frac{\scalar{b_j\t n_j,a_j\bar{\t m}_j}}{1-P_j} = 0.
\]
Upon extracting the real part of this complex equation, it comes that
\[
\sum_j \scalar{b_j\t n_j,\bar{\t x}_{j+1}} +
\sum_j \frac{\scalar{b_j\t n_j,a_j\bar{\t m}_j}}{2} = 0.
\]
Now let $A_m$, $A_n$ and $A_r$ be the areas of the triangles $(a_1\t m_1,a_2\t m_2, a_3\t m_3)$, $(b_1\t n_1,b_2\t n_2,b_3\t n_3)$ and $(\t r_1,\t r_2, \t r_3)$; in particular $A_r=A/2$ is half of the area of a unit cell. Following some elementary algebraic manipulations, the above equation remarkably turns out to be equivalent to
\[
A_m + A_n = A_r.
\]

In conclusion, in any Kagome lattice where $A_m+A_n\neq A_r$, there exist no Floquet-Bloch zero modes with $\t q$ outside of the lines $P_j=1$. This holds in particular for lattices that are not ``too distorted'' where $A_m+A_n<A_r$. In particular, weakly-distorted lattices such that all $w_j$ are non-zero are gapped at zero frequency except at the origin $\t q=\t 0$.
\section*{Appendix C. Influence of elasticity in the hinges}\label{app:hinges}
Adding rotational springs so as to account for elasticity in the hinges is arguably equivalent to adding next-nearest-neighbor interactions between nodes. Both have a stabilizing effect on the lattice and will block twisting motions leaving translations as the only periodic zero modes. When the rotational spring constants are comparable to or higher than the effective spring constants $k_j$, the Kagome lattice will be far from the regular-distorted phase transition regime of interest. Its study, from a homogenization point of view, can thus be done using standard tools as described by \citet{Hutchinson2006} or by \citet{Lubensky2015a} for instance and will not be pursued here. Instead, focus will be on lattices where the rotational spring constants are much smaller than $k_j$, specifically, where they are of second order compared to $k_j$ since such lattices will be on the brink of a phase transition.

Formally, the expansion of the motion equation~\refeq{eq:delta2ME} will change so as to include an additional second-order term $C_H'\delta^2 K_HC_H\Phi_0$ due to the presence of elasticity in the hinges. Therein, $C_H$ is the periodic compatibility matrix corresponding to the rotational springs or equivalently to the next-nearest-neighbor bonds whereas $\delta^2 K_H$ is the corresponding matrix of elastic constants.
This term is only relevant in the last step of the homogenization theory where the corrections $D'C_H'\delta^2 K_HC_H\Phi_0$ and $T'C_H'\delta^2 K_HC_H\Phi_0$ need to be added to the macroscopic motion equations. However, the former of these two corrections is zero whereas the latter is proportional to $\phi$. This is because translations remain periodic zero modes: $C_HD=0$. Accordingly, the effective constitutive law remains the same as without elasticity in the hinges up to changing the effective parameter $L$ into
\[
L = \kappa + \frac{\gamma^2}{ah}\sum_jw_j^2k_j
\]
with $\kappa = T'C_H'\delta^2 K_HC_HT/A$ and $A$ being the unit cell area.
\end{document}